\documentclass[final,5p,times]{elsarticle}
\usepackage[utf8]{inputenc}
\usepackage{graphicx}
\usepackage{enumerate}
\usepackage{setspace}
\usepackage{subfigure}
\usepackage[hyphens]{url}
\usepackage{amsmath,amsfonts,amssymb,amsthm, empheq}
\usepackage{caption}
\usepackage{wasysym} 
\usepackage{hyperref}
\usepackage{tikz}
\usepackage{multirow}
\usepackage{units}
\usepackage{adjustbox}
\usepackage{rotating}
\usepackage{hhline}
\usepackage{geometry}
\usepackage[switch, running, modulo]{lineno}
\usepackage{booktabs}
\usepackage{siunitx}
\usepackage{lscape}   

    {\clearpage\global\pdfpageattr\expandafter{\the\pdfpageattr/Rotate 90}}%
    {\clearpage\global\pdfpageattr\expandafter{\the\pdfpageattr/Rotate 0}}

\newcommand{\code}[1]{\textsf{#1}}
\newcommand{\codeSmall}[1]{\code{{#1}}}

\usepackage[lined,boxed,ruled,commentsnumbered]{algorithm2e}

\usepackage{algpseudocode}
\SetAlFnt{\footnotesize}
\SetAlCapFnt{\footnotesize}
\SetAlCapNameFnt{\footnotesize}

\SetNlSty{textsf}{}{}
\SetInd{0.1cm}{0.3cm}

\SetCommentSty{MyCommentStyle}

\makeatletter
\newcommand{\removelatexerror}{\let\@latex@error\@gobble}
\makeatother

\newcommand{\paperTitle}{
Prototyping Low-Cost Automatic Weather Stations for  Natural Disaster Monitoring}
\newcommand{\authorsCitation}{}
\hypersetup{
    breaklinks=true,
    bookmarks=true,         
    unicode=false,          
    pdftoolbar=true,        
    pdfmenubar=true,        
    pdffitwindow=false,     
    pdfstartview={FitH},    
    pdftitle={\paperTitle},    
    pdfauthor={\authorsCitation},     
    pdfsubject={\paperTitle},   
    pdfnewwindow=true,      
}

\makeatletter
\def\ps@pprintTitle{%
  \let\@oddhead\@empty
  \let\@evenhead\@empty
  \let\@oddfoot\@empty
  \let\@evenfoot\@oddfoot
}
\makeatother

\begin{document}

\begin{frontmatter}

\title{\paperTitle}

\author[unifesp]{Gabriel F. L. R. Bernardes}
\author[Cemaden]{Rogério Ishibashi}
\author[Cemaden]{André A. S. Ivo}
\author[unifesp]{Valério Rosset}
\author[unifesp]{Bruno Y. L. Kimura\corref{cor}}
\ead{bruno.kimura@unifesp.br}
\cortext[cor]{Corresponding author}

\address[unifesp]{Institute of Science and Technology, Federal University of São Paulo (ICT/UNIFESP), São José dos Campos - SP, Brazil}

\address[Cemaden]{Brazilian National Center for Monitoring and Early Warnings of Natural Disasters (Cemaden), São José dos Campos - SP, Brazil}

\begin{abstract}
Weather events put human lives at risk mostly when people might reside in areas susceptible to natural disasters. Weather monitoring is a pivotal remote sensing task that is accomplished in vulnerable areas with the support of reliable weather stations. Such stations are front-end equipment typically mounted on a fixed mast structure with a set of digital and magnetic weather sensors connected to a datalogger. While remote sensing from a number of stations is paramount, the cost of professional weather instruments is extremely high. This imposes a challenge for large-scale deployment and maintenance of weather stations for broad natural disaster monitoring. To address this problem, in this paper, we validate the hypothesis that a Low-Cost Automatic Weather Station system (LCAWS) entirely developed from commercial-off-the-shelf and open-source IoT technologies is able to provide data as reliable as a Professional Weather Station (PWS) of reference for natural disaster monitoring. To achieve data reliability, we propose an intelligent sensor calibration method to correct weather parameters.  From the experimental results of a 30-day uninterrupted observation period, we show that the results of the calibrated LCAWS sensors have no statistically significant differences with the PWS's results. Together with The Brazilian National Center for Monitoring and Early Warning of Natural Disasters (Cemaden), LCAWS has opened new opportunities towards reducing maintenance cost of its weather observational network.
\end{abstract}

\begin{keyword}
Low-cost Automatic Weather Station; Natural Disaster; Intelligent Sensor Calibration; Internet of Things.
\end{keyword}
\end{frontmatter}

\section{Introduction}
Weather monitoring is a crucial task in different domains of applications, e.g., high precision agriculture~\cite{Sawant.et.al:2017}, military missions~\cite{Winkler.et.al:2012}, outdoor entertainment and recreation~\cite{Finger.et.al:2012}, industrial production, and logistics. One of the most critical applications is natural disaster monitoring. Climate change has intensified the occurrence of natural disasters around the world~\cite{Banholzer.et.al:2014}. More intense weather events have been experienced in the last few decades, such as higher and lower temperatures, intense rains, strong winds in tropical cyclones, and intensified droughts~\cite{Aalst.and.Maarten:2006}. People might be exposed to the consequences of extreme weather events, e.g., flash flooding in underground drainage galleries, landslides on slopes, river overflow, soil and coastal erosion, infrastructural collapsing of houses and buildings. Besides human losses, climate-related natural disaster events affect the local economy by destroying productive capital, supply chains, {{and housing stock}~\cite{Boustane.et.al:2020, debortoli2017}}.  

To warn people under imminent risk of natural disasters, weather data remotely collected from weather stations play a key role. Although crowdsourcing and citizen science based {approaches~\cite{Degrossi.et.al:2014, Fava.et.al:2019,de2019}} for disaster management can obtain useful information from volunteers, the main source of accurate and reliable weather data still comes from radar and weather stations. In order to support risk management and reduce the impacts of natural disasters in Brazil, Cemaden is the government institution responsible for monitoring and providing early warnings of vulnerable areas in the whole country. {At Cemaden, weather stations are the main monitoring components in a decision-making pipeline for early warnings of natural disasters. This means that, while the weather stations themselves cannot directly monitor natural disasters, the role of such stations is essentially collecting reliable weather parameters to feed forecasting models running on Cemaden’s back-end servers. As such, the forecasting models are the entities responsible for predicting disasters and then trigger early warnings.}

\subsection{Problem Addressed and Proposed Solution}

To provide effective and efficient large-scale weather monitoring, Cemaden has an observational network composed of more than five thousand Data Collection Platforms (DCP) deployed in almost a thousand municipalities within the Brazilian territory. A DCP is implemented by a Professional Weather Station (PWS) which consists of a set of one or more sensors to measure weather events, e.g., rainfall precipitation, wind speed, wind direction, air temperature, relative humidity, and atmospheric pressure. Although a PWS provides precision and data reliability, its cost can reach tens of thousands of dollars. This implies that the deployment and maintenance of PWSs are extremely costly for large-scale weather monitoring such as Cemaden's observational network. 

As an initiative to address cost reduction, in this paper, we present a low-cost automatic weather station system, LCAWS, which we developed from commercial-off-the-shelf and open-source IoT technologies. {The key requirement of LCAWS is providing low-cost weather instrumentation that allows precision and data reliability equivalent to the PWSs employed by Cemaden to monitor natural disasters across the Brazilian territory. By reducing the costs, it is possible to execute maintenance in short periods and expand the network coverage, allowing redundancy for a failed equipment.}

\subsection{Summary of Main Contributions}

The main contributions of this work are the following:
\begin{enumerate}[(1)]
\item \textbf{Design and implementation}. We describe a detailed design of an LCAWS, including: the interaction of the embedded electronics and sensor dynamics; the data processing to produce weather parameters with the main equations; the system software architecture with the main algorithms at the automatic weather station (client node) and the cloud services (server node). 

\item  \textbf{{Weather station validation}}. We present a consistent experimental methodology to validate LCAWS which includes three main steps. First, the weather station deployment together with a PWS of reference. Second, the data acquisition with the same frequency within a one-month continuous period of data sampling from measurements of atmospheric pressure, air temperature, relative humidity, rain precipitation, wind speed, and wind direction. Third, the data analysis to compare the weather parameters produced by the stations from different performance indicators. 

\item \textbf{{Intelligent sensor calibration and data correction}}. We apply a robust methodology for calibrating the sensors in order to correct the weather parameters by means of linear and machine learning regression models. In such a methodology, for each weather sensor, we select the best models from a broad set of candidate regression models. To do so, for each candidate model, we ran experiments exhaustively with different randomized train-validation and test datasets by using a $k$-fold cross-validation based machine learning pipeline. When applying the best models to correct the LCAWS weather parameters, they allowed an improvement on R-squared coefficient determination (R2) from 0.93 to 0.99 for digital sensors, and from 0.31 to 0.97 for magnetic sensors. In other words, this means a root mean squared error (RMSE) reduction of up to 70\% for digital sensors and and up to 80\% for magnetic sensors. Such an intelligent approach allowed an improvement in such a way that there were no significant differences in the weather parameters produced by the LCAWS and PWS, i.e., T-Test's \textit{p}-value greater than 0.05.

\item \textbf{{A good candidate solution to reduce costs in natural disaster monitoring}}. Using a robust methodology, we showed that an LCAWS of a few hundred dollars has the potential to provide weather data as reliable as a PWS applied for natural disaster monitoring. When validating a reliable LCAWS prototype, new opportunities have been presented together with Cemaden to expand its national weather observational network and reduce the cost of maintenance of its DCPs. 
\end{enumerate}

\subsection{Related Work}

{In line with the hardware architecture proposed in this paper, some Arduino-based low-cost prototypes especially designed for environmental monitoring are also present in the literature \cite{sabharwal2014low, Saini.et.al:2016, Lockridge.et.al:2016, Strigaro.et.al:2019}. Among them, ground weather 
 {stations developed by }
\citet{sabharwal2014low} and \citet{Saini.et.al:2016} are based on the combination of Arduino Uno and ZigBee technologies. 
 {Whereas \citet{Lockridge.et.al:2016} designed} a Sonde for monitoring marine 
 {environments, \citet{Strigaro.et.al:2019}}
also demonstrated that low-cost weather stations based on COTS IoT devices are accessible solutions able to produce data of appropriate quality for natural resource and risk management. }

\citet{Benghanem:2009} proposed a wireless data acquisition system (WDAS) and a low-cost weather station, whose hardware architecture was based on PIC (Parallel Interface Controller) 16F877 microcontroller and a communication module using RF Monolithics TX5002 and RX5002. Likewise, \citet{Tenzin.et.al:2017} developed a low-cost weather station for a smart agriculture application based on PIC24FJ64 microcontroller and a Xbee module, comparing results with a commercial and more costly station.

\citet{Shaout.et.al:2014} present a low-cost weather prototype for measuring wind speed, wind directions, and temperature. These parameters are used in combination with a neural network in order to estimate a dressing index, i.e., the number of clothes a person should use under certain conditions. The hardware architecture consists of a Freescale HCS12 family's Dragon12-Plus2 and a MC9S12DG256 microcontroller (16-bit CPU, 256 KB flash memory, \unit[12]{KB} RAM, \unit[4]{KB} EEPROM). \citet{Kodali.and.Mandal:2016} explore the capabilities of ESP8266 (\unit[4]{MB} RAM, \unit[128]{KB} ROM) to prototype a very low cost weather station to acquire temperature, pressure, light, and rain drop sensors.

Different from these related works, in our study, we focus on the context of the natural disaster monitoring system which presents stronger operation requirements and high accuracy of gathered data. We validate an LCAWS by comparing results with a PWS of reference for natural disaster monitoring, applying a consistent experimental methodology in order to fulfill weather monitoring requirements of Cemaden---a national monitoring center. In order to improve the accuracy of collected weather data, we propose and implement as an intelligent calibration system based on linear and machine learning regression models, which is a contribution that does not appear in related works.

The remainder of this paper is organized as follows: Section~\ref{sec:Cemaden} presents Cemaden's remarks on the problem we addressed in this paper. Section~\ref{sec:LCAWS-prototype} describes our LCAWS prototype. Section~\ref{sec:LCAWS-dataprocessing} describes the weather data processing for LCAWS. Section \ref{sec:LCAWS-validation} describes the weather station validation.  Section~\ref{sec:LCAWS-calibration} presents the intelligent sensor calibration and data correction approach we proposed. Finally, Section~\ref{sec:Conclussions} brings our main conclusions, as well as the LCAWS's limitations to be addressed in future work.

\section{The Brazilian National Center for Monitoring and Early Warning of Natural~Disasters}
\label{sec:Cemaden}

\begin{figure*}[ht]
    \centering
    \includegraphics[scale = 0.65, trim = 0.5cm 0.5cm 0cm 0.5cm]{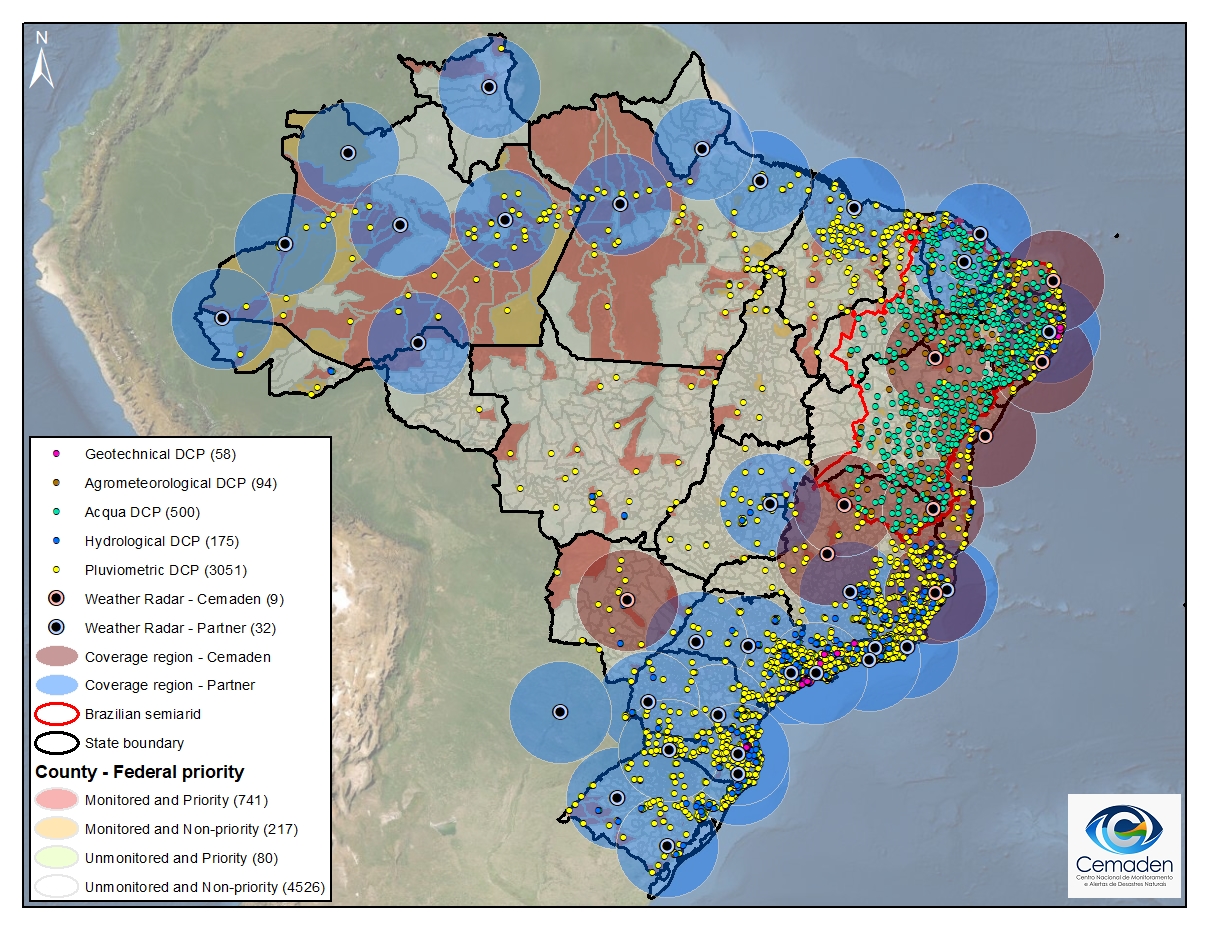}
    \caption{Distribution of Cemaden's observational network.}
    \label{fig:RO-Cemaden}
\end{figure*}

Cemaden is a national center created in 2011 under the Ministry of Science, Technology, and Innovation of the Brazilian government. Besides providing uninterrupted natural disasters monitoring and early warnings, Cemaden also carries out research and technology innovation to contribute to its monitoring systems in order to reduce the number of victims living in risk areas around the country~\cite{decretoCemaden2011}. Nowadays, Cemaden is monitoring 958 municipalities in the entire Brazilian territory, mostly for geological (landslides) and hydrological (floods) events often caused by intense or sparse rain events~\cite{Mendes2018}. To reach its goals, Cemaden holds an observational network, currently consisting of 5857 different kinds of DCPs and nine weather radars. Such a network covers metropolitan areas subject to natural disasters while gathering meteorological data in real-time. Cemaden also has numerous partnership agreements to provide and receive weather information from other observational networks (radars and DCPs), expanding its network for better coverage and redundancy. The coverage map in Figure~\ref{fig:RO-Cemaden} shows the national territory with the priority municipalities overlaid by the distribution of the observational systems.

\subsection{DCP Types for Natural Disaster Monitoring}

Every Cemaden DCP is a Professional Weather Station (PWS) equipped with a tipping bucket rain gauge, and able to accommodate additional sensors of specific groups of natural disaster monitoring. Such groups are divided into five DCP/PWS categories:{Pluviometric}, {Hydrological}, {Geotechnical}, {Agrometeorological}, and {Acqua}.

A {pluviometric} DCP is used for monitoring natural disasters in general, which are often triggered by rain, and it is usually composed of only a rain gauge. A {hydrological} DCP is used for flood monitoring from a rain gauge, a level radar for river level measurement, and a photography camera. A {geotechnical} DCP consists of a rain gauge and six soil humidity sensors (up to \unit[3]{m}) to be used together with a robotic total station to help in the landslide monitoring. An {agrometeorological} DCP is used for drought monitoring in the semiarid region of Brazil. It consists of a rain gauge, a thermo hygrometer for air temperature and relative humidity, an anemometer for wind speed, a vane for wind direction, four soil temperature and humidity sensors, a pyranometer, and a net radiometer for solar radiation measurements. An {acqua} DCP is also used for drought monitoring in the Brazilian semiarid region from a rain gauge and two superficial soil humidity sensors. 

In spite of its importance for governmental decision-making, the expansion and maintenance of almost six thousands weather instruments in a territory of \unit[8,516,000]{km$^2$} impose financial and technological challenges and logistics and communication difficulties and vulnerabilities.

\subsection{Frequent Vulnerabilities of DCPs}

Based on historical information and maintenance reports on Cemaden's DCP network, it was possible to identify the most frequent vulnerabilities that cause service interruption, classify them into types, and determine their impact from the percentage of affected devices. Table~\ref{tab:affect} presents an analysis of vulnerabilities regarding a 2017 survey.

\begin{table}[h]
    \centering
    \footnotesize
    \caption{Frequent vulnerabilities found in Ceamaden's DCP network.}
    \label{tab:affect}
\begin{tabular}{l l l}
    \toprule
    \textbf{Vulnerability} & \textbf{Type} & \textbf{Affected Devices} \\
    \midrule
        Rain gauge clogging         	& Physical 		& {45.60}{\%} \\
        Theft, vandalism, and damage 	& Human  		& {5.37}{\%} \\ 
        Battery                     					& Technological	& {4.49}{\%} \\ 
        Communication failure       			& Technological	& {12.05}{\%} \\ 
        No internet coverage         			& Technological	& --    \\ \bottomrule
    \end{tabular}    
\end{table}

{Rain gauge clogging} is a physical vulnerability that represents the greatest number of occurrences, a total of 35,185 occurrences in 2017, which affected 2671 devices. In most cases, the problem of funnel clogging is caused by dirt, falling leaves and small insects that interrupt water flow through the siphon of the tipping bucket rain gauge.
 
 {Theft, vandalism, and damage} are human type vulnerabilities that affected around
  5.37{\%} of devices in 2017. In particular, there are many goats raised for milk and meat in the semiarid region. These animals have a peculiar diet, and it is not unusual to gnaw the DCP's cables, sensors, and even the holding box. 
 
 Regarding the technological vulnerabilities, we highlight that the most impactful ones are battery maintenance, communication failures, and the absence of internet coverage access. A {battery} is a type of commodity item that has a short life cycle and needs to be replaced every two~years. Thus, it is important to have a good schedule for battery replacement in maintenance campaigns. {Communication failures} were registered in 706 devices by data transmission problems with a total of 18,750 cases in 2017. Currently, the DCP communication is based on GSM/GPRS technologies that uses the mobile telecommunication infrastructure. Thus, there are at least two factors that could degrade signal quality and cause a failure: long distances between the telecommunication tower and DCP, and obstructions between them, e.g., mountains, hills, buildings, even a strong rain can attenuate radio signals. {No internet coverage} is an extension of communication failures. This technological vulnerability represents a limitation to expand or rearrange the observational network, as the telecommunication towers typically cover the most populated areas and the coastal zone of the country. It means that there are difficulties to install instruments in rural areas (e.g., to monitor drought), top of mountains, and even other basins where rain could contribute to a natural disaster in a risk area. Nowadays, Cemaden has not deployed any DCPs in areas with no telecommunication coverage.

\subsection{Maintenance Schedules: A Cost-Related Challenge to Mitigate {the Impacts of}~Vulnerabilities of~DCPs}

The aforementioned vulnerabilities can directly affect the weather condition analysis at Cemaden's Situational Room and, as consequence, affect the quality of the warning messages sent to Civil Defenses. In order to mitigate {the impacts of} these vulnerabilities, an effective preventive maintenance schedule in short periods is a pivotal measure. 
However, Cemaden is a governmental center with very limited financial and human resources. Currently, it is possible to execute a device's maintenance every eighteen months. 

In this context, one of its greatest challenges to mitigate the {impacts of} vulnerabilities in its observational network is to execute maintenance within appropriate periods (less than six months) using the same human resources and available budget. From the vulnerability analysis, we observed that they are not linked to the robustness of hardware, but to the applied technology. Cemaden's DCPs use expensive imported components (e.g., sensors, power regulators, solar panels, AC chargers, batteries, dataloggers, network interface controllers, enclosures, among others), which makes maintenance expensive, either preventively or for replacing failed components. {Indeed, low-cost stations would be exposed to the same vulnerabilities in the field. However, an LCAWS with similar functionalities and measurement accuracy should be a good cost alternative to a PWS, allowing the execution of maintenance in shorter periods and also an expansion of the network coverage, giving redundancy for a failed equipment.}

The maintenance regards other preventive actions, such as the installation of fences around most exposed equipment, partnerships with Civil Defenses for maintenance, informative boards, etc. Communication failures and the absence of internet access coverage are vulnerabilities that require efforts also from third parties such as mobile network operators. To improve connectivity beyond the access provided by network operators, e.g., by deploying an ad-hoc network to connect DCPs and Cemaden's servers, the DCPs have to be updated. However, besides the need for acquiring new components (i.e., incurring cost of imported components), improving communication capacity would be limited to the DCP manufacturer's solutions.

\subsection{LCAWS as a Key Enabler for Low-Cost Reliable Weather Monitoring}

As discussed above, the cost reduction in the maintenance process for a reliable weather monitoring is a key challenge for Cemaden, regarding its limited financial and human resources. In this context, prototyping Low-Cost Automatic Weather Stations (LCAWS) can represent a great opportunity to reduce the maintenance cost composition, since it uses easily found Commercial Off The Shelf (COTS) components, representing just a fraction of the values of actual DCP/PWS components. Besides a significant cost reduction in extreme scales ($\approx $1/30) without losing autonomy or functionalities, the openness feature from open-source IoT technologies employed by LCAWS leverages other benefits, e.g., the flexibility for updating hardware/software components without being confined to single manufactures' solutions; the possibility of acting in different domains of research and development, such as low-cost technologies for environmental sensors, sensor data communication, data loggers, among others. In the next section, we describe the design and implementation of our LCAWS prototype.

\section{The Prototype of a Low-Cost Automatic Weather Station}
\label{sec:LCAWS-prototype}

The LCAWS presented in this paper is composed of six main components mounted on a structure of vertical and horizontal axes, as seen in Figure~\ref{fig:LCAWS-components}a and described in \mbox{Table~\ref{tab:LCAWS-components}}. The weather measurements are provided by magnetic and digital sensors. The magnetic sensors are the wind vane, cup anemometer, and tipping bucket rain gauge, which are implemented by the components 1, 2, and 3, respectively. The sensors for measuring atmospheric pressure, air temperature, and relative humidity are implemented by a single digital module that is housed in component 5. Component~4 is a photo-voltaic solar panel responsible for charging a \unit[12]{V} \unit[6]{Ah} lead-acid battery which powers the weather station. The battery is housed in component 6, the datalogger, which is responsible for acquiring, processing, storing, and transmitting the weather measurements. In the next subsections, we present details on the magnetic and digital sensors, the datalogger components, as well as the software architecture of the system.

\newcommand\circleTag[1]{%
  \tikz[baseline=(X.base)]
  \node (X) [draw, shape=circle, inner sep=0.9pt] { \textbf{\textsf{#1}}};
}

\newcommand{\figLCAWSComponents}{
    \begin{minipage}[c][10.2cm][t]{5.3cm}
        \centering
        \subfigure[][Weather station components]{
            \includegraphics[scale=0.6, trim=0cm 0.5cm 0cm 0.3cm]{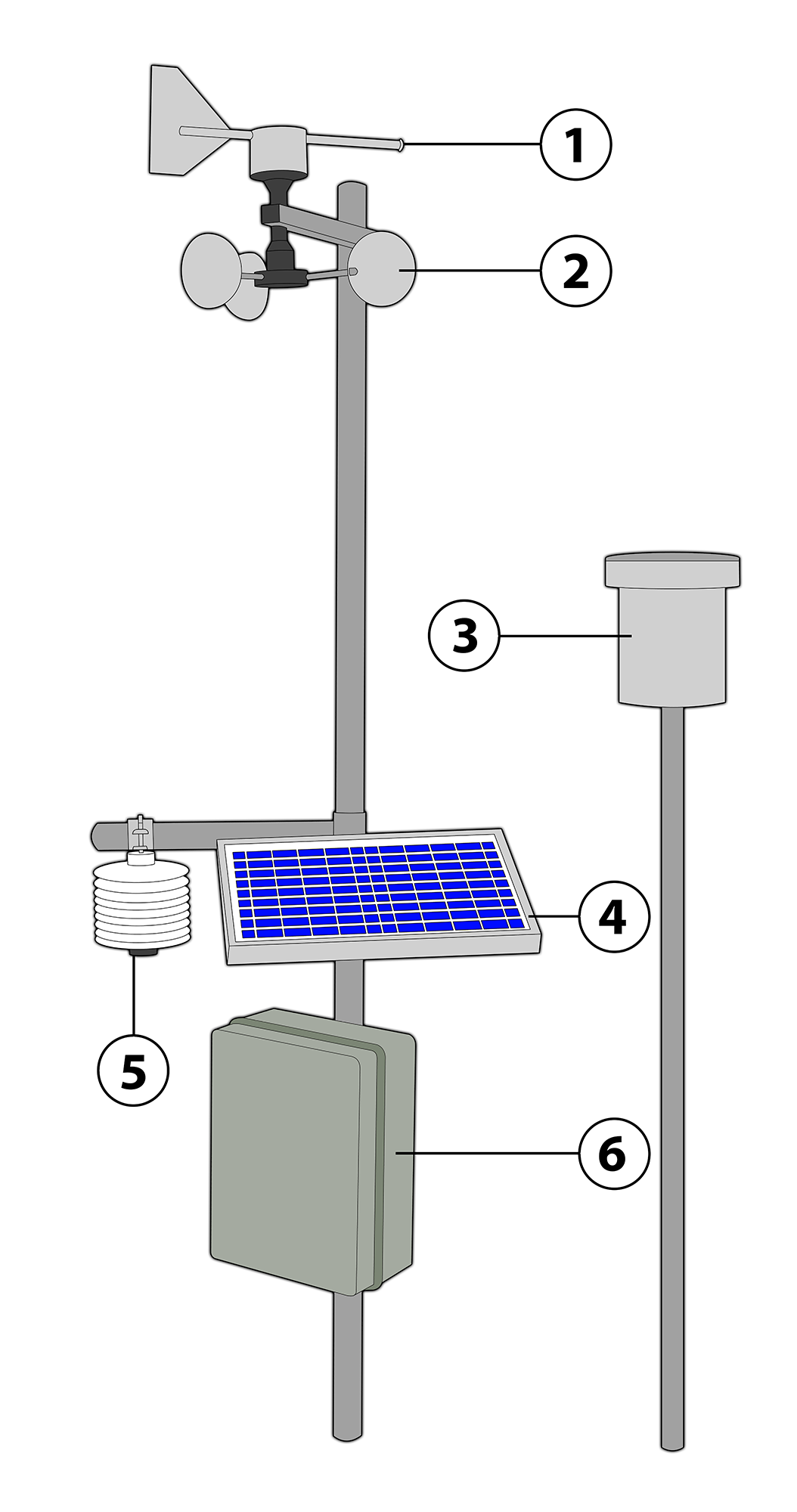}
        }
    \end{minipage}
}

\newcommand{\figLCAWSDatalogger}{
    \begin{minipage}[c][5.5cm][t]{4.5cm}
        \centering
        \subfigure[][Components inside datalogger]{
            \includegraphics[scale=0.13, trim=0cm -0.3cm 0cm 0.0cm]{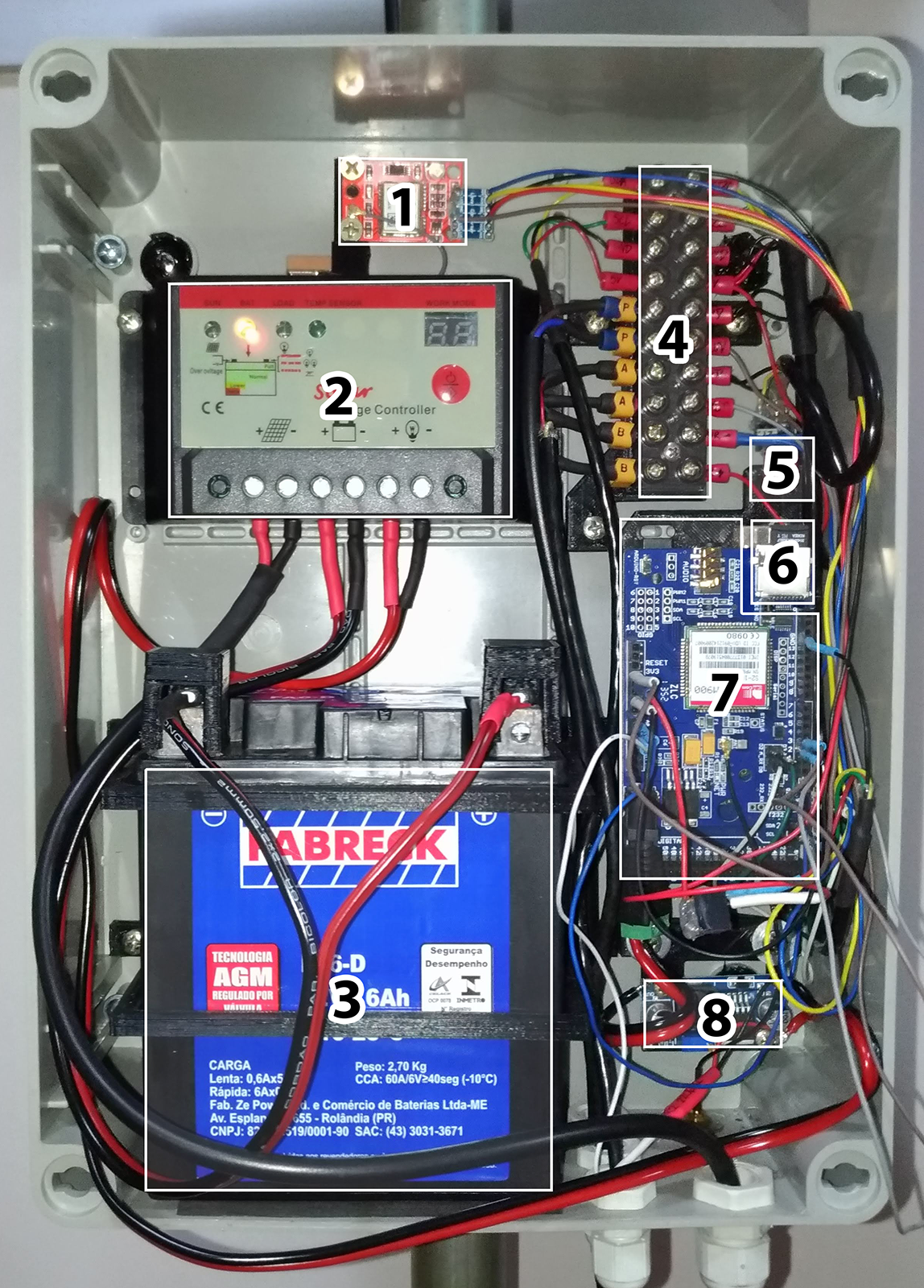}
        }
    \end{minipage}
}
  
\newcommand{\tabLCAWSComponents}{%
    \centering
    \captionof{table}{LCAWS components.}
    \label{tab:LCAWS-components}
    \scriptsize
    \renewcommand{\arraystretch}{1.2}
    \addtolength{\tabcolsep}{-4pt} 
    \begin{tabular}{@{}l@{} p{1.8cm} p{4.5cm}}
        \toprule
        & {Component} & {Description} \\
        \midrule
            \circleTag{1} & {Wind vane:} & 
                Wind vane magnetic sensor (reed switch) of $\ang{45}$ resolution (N, NE, E, SE, S, SW, W, NW). \\
            \circleTag{2} & {Anemometer:}
                &  Wind speed magnetic sensor (reed switch) of one pulse per full rotation of \unit[75]{mm} diameter aluminum cups, \unit[147]{mm} radius to cup center. \\
            \circleTag{3} & {Rain gauge:} 
                & Tipping bucket rain gauge with magnetic sensor (reed switch) of \unit[0.25]{mm} precipitation per pulse. \unit[150]{mm} diameter collector. \\
            \circleTag{4} & {Solar panel:} & Yingli YL055P-17b \unit[55]{W} peak power solar panel used to charge the battery. \\
            \circleTag{5} & {Sensor housing:}  
                & Bosch BME280 combined digital sensors of air temperature, atmospheric pressure, and relative humidity.\\
            \circleTag{6} & {Datalogger}: & \\
        ~.1 & GPS receiver:       & u-blox Neo-6M GPS receiver module.\\
        ~.2 & Charge controller:  & Generic \unit[12]{V}/\unit[24]{V} \unit[10]{A} Charge controller.\\
        ~.3 & Battery:            & \unit[12]{V} \unit[6]{Ah} AGM lead-acid battery. \\
        ~.4 & Terminal block:     & Generic terminal block for cable                                   connections.\\
        ~.5 & Internal sensors:   & Bosch BME280.\\
        ~.6 & Storage:            & \unit[2]{GB} MicroSD card.\\
        ~.7 & Processing~\&      & Arduino Mega 2560, \\
           & Telemetry:     & Epalsite GPRS Shield V1.0 (SIM900).\\
            
        ~.8 & Voltage regulator:  & DC-DC step-down voltage regulator.
        \\ \bottomrule
    \end{tabular}
}

\begin{figure*}[ht]
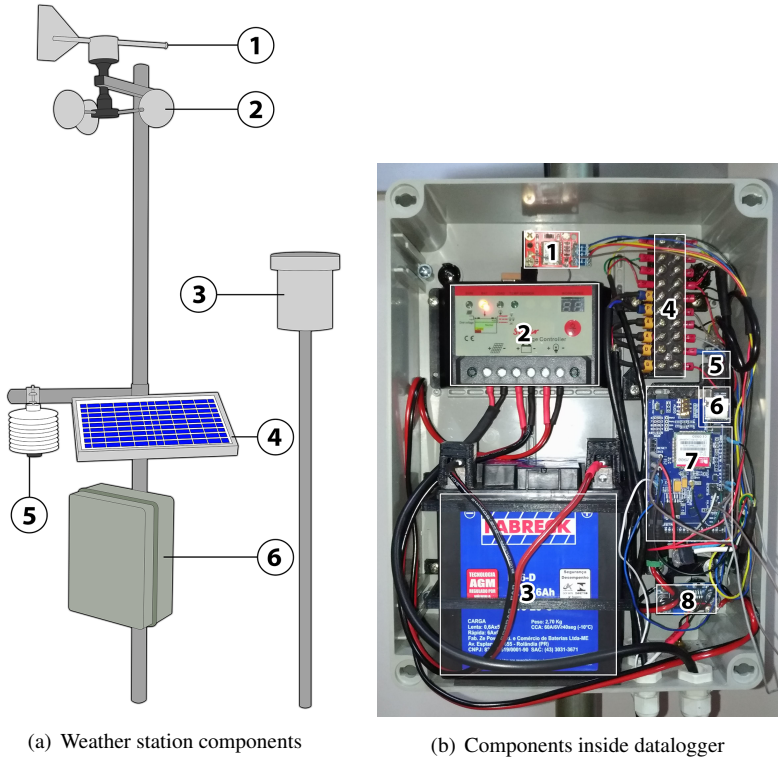

    \begin{minipage}[c][10.7cm][t]{0.6\linewidth}
        \figLCAWSComponents
        \figLCAWSDatalogger
        \captionof{figure}{Components of the Low-Cost Automatic Weather Station (LCAWS).}
        \label{fig:LCAWS-components}
    \end{minipage}
    \begin{minipage}[c][8.5cm][t]{0.4\linewidth}
        \tabLCAWSComponents
    \end{minipage} 
\end{figure*}

\subsection{Magnetic Sensors}

The wind vane, cup anemometer, and tipping bucket rain gauge are magnetic sensors. Each consists of one or more reed switches and a permanent magnet that is attached to a particular mechanical device. Reed switches are devices that open or close an electrical circuit when influenced by a magnetic field. All reed switches in these specific sensors are of the normally open type, i.e., they remain open by default until a nearby magnetic field causes them to close. Each mechanical device has a specialized structure designed to interact physically with the environment so that movement of its parts causes a permanent magnet to influence one or more reed switches. By causing a reed switch to allow electricity to pass, the mechanical device's movement generates electric signals which can consistently and reliably be interpreted as a specific weather phenomenon.

Each weather device produces particular movement dynamics. Air flow causes the rotary part of the wind vane to rotate on a vertical axis until it points to the direction from where the wind is blowing. The permanent magnet fixed to the rotary part moves on top of eight reed switches equally spaced by $\ang{{45}}$ fixed to the static part of the wind vane so that only one reed switch is on at any given time. Each of the reed switches is in series with a resistor of a distinct value varying uniformly from \unit[10,000]{$\Omega$} to \unit[80,000]{$\Omega$}. When that particular circuit is closed, the resistance of the overall circuit is uniquely related to a particular wind direction. The overall circuit is in series with a fixed reference resistor of \unit[4700]{$\Omega$}, forming a voltage divider. The wind direction is thus associated with the voltage measured in the wind vane circuit. 

The mobile component of the anemometer is composed of a permanent magnet and three \unit[75]{mm} diameter aluminum cups mounted on horizontal arms. Air flow makes this mobile part rotate on a vertical axis and once per full rotation the permanent magnet triggers a reed switch which is fixed to the static part. These rotations generate electric pulses that, along with a timestamp associated with each pulse, can be used to determine wind speed.

The tipping bucket rain gauge consists of a \unit[150]{mm} diameter funnel where rainwater collects and drips down to a tipping bucket. The tipping buckets are the only moving parts and store water in one of two buckets. When one bucket accumulates a specific volume of water, it tips over, emptying itself and exposing the other one. Once that other bucket is full, it tips over and repeats the process. A permanent magnet placed between the two buckets triggers one of two reed switches, creating a pulse that is interpreted as a precise amount of precipitation that has occurred since the last pulse.

\subsection{Digital Sensors}

{Component~5 is a sensor housing which accommodates and protects three digital sensors: air temperature, atmospheric pressure, and relative humidity. Such sensors are combined into a single Bosch BME280~\cite{BME280.datasheet:2018} module. While providing low power consumption (\unit[3.6]{$\mu$A} at \unit[1]{Hz} measurement), high accuracy and resolution, the BME280 module encapsulates the sensors into an electronic component of small dimensions (11.5 $\times$ 15 mm). }

Different settings of oversampling and filtering allow tailoring data rate, noise reduction, response time, and energy consumption. An impulse response filter (IIR) can remove short fluctuations in data sampling. The small dimensions and versatile features allow sensor implementation in small devices such as handsets, GPS modules, watches, in different applications, e.g., home automation, indoor navigation, fitness, and GPS refinement. Differently, in this paper, we apply the BME280 to implement an automatic weather station.

\subsection{Datalogger}

Component~6 is the datalogger, which  is  a  key  instrument  that  contains all the circuitry  responsible for sensor  data  acquisition,  processing, and  storage,  for  telemetry,  and  for powering  the  weather station. The datalogger is composed of eight main internal components, as seen on Figure~\ref{fig:LCAWS-components}b: 

\begin{enumerate}[(.1)]
    \item {{GPS receiver}}. 
Aside from useful to synchronize time and localization, the GPS receiver provides a source of data for further investigations, e.g., correlation between signal-to-noise ratio (SNR) from different satellites and the weather measurements.
    
    \item {{Charge controller}}. It is a device that regulates the highly variable voltage and current supplied by a solar panel and charges a battery following a charge curve. The charge controller used in our prototype is a generic \unit[10]{A} max \unit[12/24]{V} pulse width modulation (PWM) controller.
    
    \item {{Battery}}. The weather station is powered by a \unit[12]{V} \unit[6]{Ah} lead-acid absorbed glass mat (AGM) motorcycle battery.
    
    \item {{Terminal block}}. A 10-pin terminal block (component 6.4) is used to facilitate maintenance and secure stable electrical connections between sensors and the microcontroller.
    
    \item {{Internal weather sensors}}. A Bosch BME280 sensor module is installed inside the datalogger housing to monitor internal temperature and humidity.
    
    \item {{Storage}}. Data is stored in a \unit[2]{GB} microSD card.
    
    \item {{Processing and telemetry}}.  The Arduino Mega 2560 is the microcontroller used to acquire, process, store, and transmit data from sensors. Coupled with a SIM900 GPRS shield, it sends the weather data automatically to a cloud remote server.
    
    \item {{Voltage regulator}}. An LM2596S DC-DC step-down voltage regulator chip, which converts \unit[10--14]{V} inputs from the battery to a fixed output at \unit[7]{V} to the Arduino.
\end{enumerate}

\subsection{System Software Architecture}

The LCAWS system software is based on a client--server architecture, as shown in Figure~\ref{fig:LCAWS-sysarch}. The automatic weather station~(AWS) is composed of a client program and routines deployed on the datalogger. On the server side, the cloud services~(CS) are provided by programs and routines for processing and data corrections deployed on a cloud. The rationale of AWS and CS operations are summarized in Algorithms~\ref{alg:AWS} and~\ref{alg:CS}, respectively.

\newcommand{\algLWCS}{%
    \singlespacing
    \begingroup
    \removelatexerror
        \begin{algorithm}[H]
            \label{alg:AWS}
            \SetKwProg{proc}{procedure}{:}{}
            \SetKwProg{prog}{program}{:}{}
            \DontPrintSemicolon
            \caption{AWS operation.}
            \KwIn{\\
            ~~~~$N_\tau$ = 10, tuples $\tau$ per file.\\
            ~~~~$t_\code{s}$ = \unit[1]{min}, a sampling standby.\\
            ~~~~server = IP-port,  address info.}
            \prog{\upshape{\code{client($N_\tau$, $t_\text{s}$, \textrm{server})}}}
            {
                \code{init\_modules()}\;
                $S$ = \code{init\_sensors()}\;
                \While {\upshape{\texttt{true}}}
  	            {   
  	                \code{$f$ = create\_file()}\;
                    \ForEach {\upshape{\text{1 to $N_\tau$}}}
  	                {
  	                    \code{$\tau$ = get\_sensordata($S$)}\;
                        \code{write\_file($f$, $\tau$)})\;
                        \code{standby($t_\text{s}$)}\;
                    }
		            \code{send\_to(\textrm{server}, $f$)}\;
		            \code{close($f$)}\;
                }
            }
            \proc{\upshape{\code{get\_sensordata($S$)}}}
            {   
                \code{$\tau$ = \texttt{null}}\;
                \ForEach {$s \in S$}
                {
                    \code{$v$ = read\_sensor($s$)}\;
                    \code{$\tau$ = cat($\tau$, $v$)}\;
                }
                \code{$t_\text{ts}$ = get\_timestamp()}\;
                \code{$\tau$ = cat($\tau$, $t_\text{ts}$)}\;
                \KwRet $\tau$\;
            }
\end{algorithm}
\endgroup
}

\newcommand{\algCloud}{%
    \singlespacing
    \begingroup
    \removelatexerror
        \begin{algorithm}[H]
            \label{alg:CS}
            \SetKwProg{proc}{procedure}{:}{}
            \SetKwProg{prog}{program}{:}{}
            \DontPrintSemicolon
            \caption{CS operation.}
            \KwIn{\\
            ~~~~~~\code{DB}$_L$ = \texttt{null}, an initialized $\tau$ database.\\
            ~~~~~~$\overline{\code{DB}}_L$ = \texttt{null},  an initialized $\overline{\tau}$ database. \\
            ~~~~~~$t$, UTC starting time for data processing. \\
            ~~~~~~$i$ = \unit[1]{hour}, time interval from $i$. }
            \prog{\upshape{\code{server()}}}
            {
                \While {\upshape{\texttt{true}}}
  	            {   
  	                \code{$f$ = create\_file()}\;
		            \code{recv\_from(\textrm{client}, $f$)}\;
		            \code{append(DB$_L$, $f$)}\;
		            \code{close($f$)}\;
                }
            }
            \prog{\upshape{\code{process\_data(DB$_L$, $\overline{\code{DB}}_L$, $i$, $g$)}}}
            {
                \While{\upshape{\textbf{not} \code{end\_of(DB$_L$)}}}
                {
                    \code{$d$ = subset(DB$_L$, $t \leq t_\text{ts} < t + i$)}\;
                    $\overline{\code{AP}}$ = \code{summary($d_\text{AP}$)}
		    	       \tcp*[r]{Equation~(\ref{eq:mean_BME280})} 
	    		    $\overline{\code{AT}}$ = \code{summary($d_\text{AT}$)}
	    		        \tcp*[r]{Equation~(\ref{eq:mean_BME280})}
			        $\overline{\code{RH}}$ =     \code{summary($d_\text{RH}$)}
			            \tcp*[r]{Equation~(\ref{eq:mean_BME280})}
			        $\overline{\code{RG}}$ = \code{summary($d_\text{RG}$)}
			            \tcp*[r]{Equation~(\ref{eq:mean_P})}
			        $\overline{\code{WS}}$ = \code{summary($d_\text{WS}$)}
			            \tcp*[r]{Equation~(\ref{eq:mean_WS})}
			        $\overline{\code{WD}}$ = \code{summary($d_\text{WD}$)}
			            \tcp*[r]{Equation~(\ref{eq:mean_WD})}
			    
			        $\overline{\tau}$ = \code{cat($\overline{\code{AP}}$, $\overline{\code{T}}$, $\overline{\code{RH}}$, $\overline{\code{P}}$, $\overline{\code{WS}}$, $\overline{\code{WD}}$)}\;
			    
                    \code{append($\overline{\code{DB}}_L$, $\overline{\tau}$)}\;
                    $t$ = $t + i$\;
                }
            }  
\end{algorithm}
\endgroup
}


 

\newcommand{\figSysArch}{%
    \centering
    \includegraphics[scale=0.7, trim=0cm 0cm 0cm 0cm]{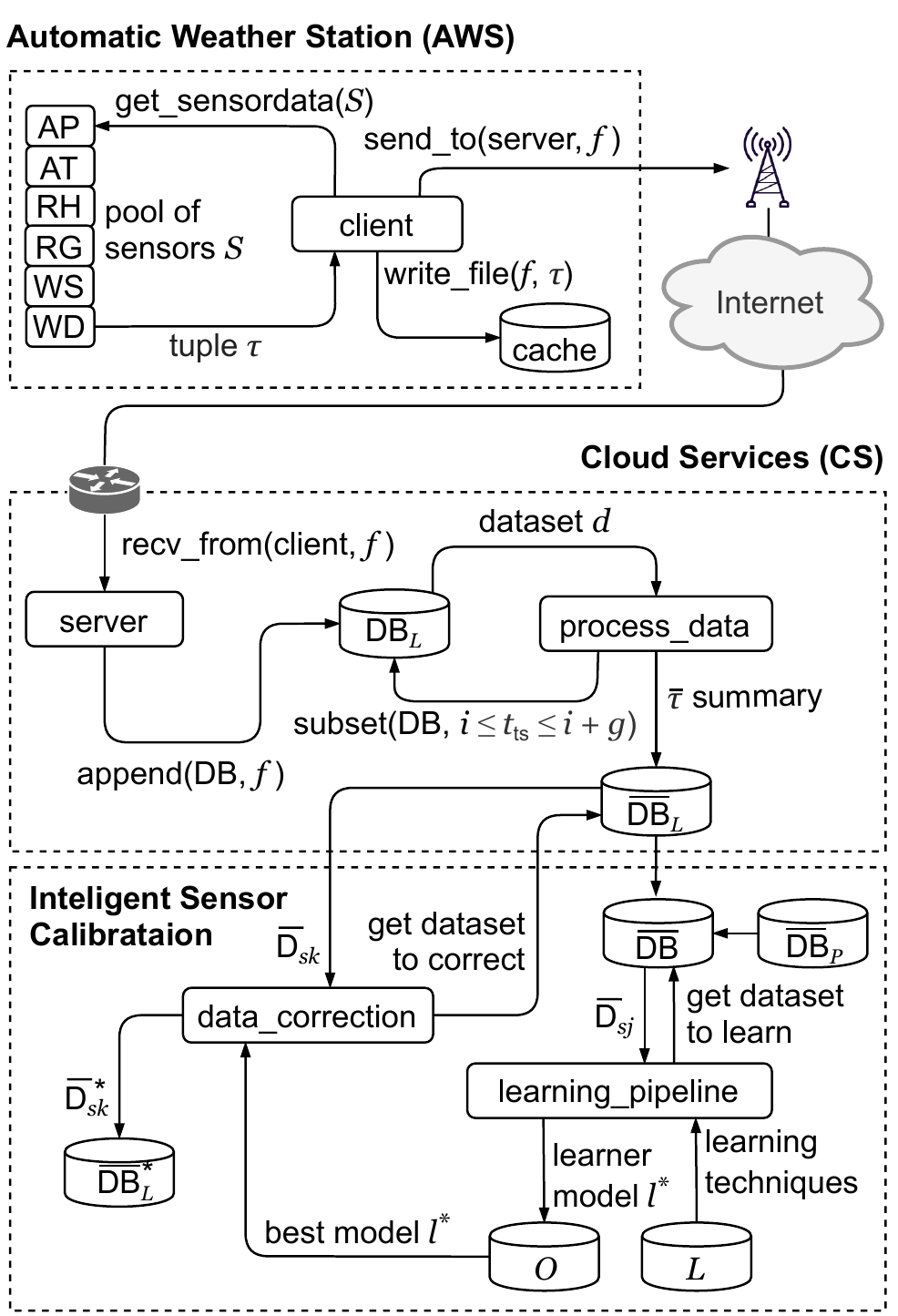}
    \caption{LCAWS system software architecture.}
    \label{fig:LCAWS-sysarch}
}

\begin{figure*}[h]
    \begin{minipage}[c][10.7cm][t]{7cm}
        \figSysArch
    \end{minipage}
    \hspace{0.1cm}
    \begin{minipage}[c][10.7cm][t]{4.5cm}
        \algLWCS
    \end{minipage}
    \hspace{0.2cm}
    \begin{minipage}[c][10.7cm][t]{6.1cm}
        \algCloud
    \end{minipage} 
\end{figure*}

The client program is implemented with the C++ Arduino API~\cite{Arduino.referece} in order to run on the microcontroller deployed on the weather station. The weather sensors are gathered into a pool of $S$ sensors: atmospheric pressure~(AP), air temperature~(AT), relative humidity~(RH), rain gauge~(RG), wind speed~(WS), and wind direction~(WD). The client program runs continuously and collects the sensor data once a minute, hence there is a standby time $t_\code{s}$ between each data acquisition cycle. The generic procedure \codeSmall{   {get\_sensordata()}} reads the sensors and returns the following 7-tuple:
\begin{equation}
    \tau = \{\text{AP}, \text{AT}, \text{RH}, \text{RG}, \text{WS}, \text{WD}, t_\code{ts}\}
\end{equation}
where $t_\code{ts}$ is the UTC timestamp of each data acquisition cycle on the pool of sensors. Inside the \codeSmall{   {get\_sensordata()}} procedure, the \codeSmall{   {read\_sensor()}} procedure reads out the value available for each sensor~$s$.

A file $f$ is created to store a dataset of $N_\tau = 10$~tuples~$\tau$ in a cache on the local file system. The file $f$ is sent to the server via TCP/IP over the mobile operator network under configurable data frequencies. For the sake of real-time weather monitoring, the periodic communication can be accomplished at the rate of the data acquisition from the sensors, i.e., the sending of sensor data tuple every minute cycle as soon as it is measured. {This sampling frequency should be enough, since it is higher than the one configured in Cemaden’s DCPs on the field, which is one sample every 10 min to monitor natural disasters. If the internet connection is completely interrupted, the data continue being buffered in the local files and is transmitted when the internet connection is reestablished. If a connection is unstable and packet drop occurs, the TCP connection between the AWS client and the CS server is responsible for providing the reliable transmission required to consistently send the weather data cached in local files.} In our proof-of-concept prototype for automatic data acquisition, the data transmission is implemented with a GPRS shield connected to the Arduino microcontroller. Nevertheless, once the LCAWS design is a COTS-based modular architecture, there is flexibility to apply other up to date deployable long-range IoT data transmission technologies~\cite{Centenaro.et.al:2016, Lauridsen.et.al:2017}, e.g., { LoRA~\cite{augustin2016study}, SigFox~\cite{zuniga2016sigfox}, and NB-IoT~\cite{schlienz2016narrowband}}.

On the remote side, cloud services are provided from a server program and a program for data processing. The \codeSmall{ {server}} is implemented in Python and runs continuously in order to receive the file $f$ and append $N_\tau$ tuples of raw data values into a local database \codeSmall{ {DB}}. Asynchronously, \codeSmall{ {process\_data}} is another program implemented in R Language, which processes weather measurements in raw values and produces the corresponding weather parameters. To do so, a subset of raw data $d$ in $\codeSmall{ {DB}}_L$ is selected regarding $t_\code{ts}$, which is the measurement UTC timestamp. Since the weather parameters are usually summarized in statistics per hour, the dataset~$d$ considers an interval $[t, t+i-1]$, where~$t$ is a given UTC starting time, and \unit[$i = 60$]{min} is the dataset time interval. For each 1 h subset $d$, we determine the weather parameter for each sensor $s \in S$ from the procedure \codeSmall{ {summary()}}. As highlighted in Algorithm~\ref{alg:CS}, such a procedure applies equations to give the parameters across the different types of weather data. These equations are discussed in the next section.

{As a service CS at the server side, we have the intelligent sensor calibration programs. The processed LCAWS weather data stored in $\overline{\code{DB}}_{L}$ and the ones in $\overline{\code{DB}}_{P}$ are merged into a general database $\overline{\code{DB}}$. The database $\overline{\code{DB}}_{P}$ stores the corresponding weather parameters obtained from a professional weather station of reference. As such, a dataset $\overline{\code{D}}_{sj} \in \overline{\code{DB}}$ of a sensor $s \in S$ consists of the weather parameters of both low-cost and professional weather stations, regarding a $j$ period of observation. Given $\overline{\code{D}}_{sj}$ and a set $L$ of candidate machine learning-based regression techniques, the procedure \codeSmall{ {learning\_pipeline}} is responsible for constructing the learner model $l^{*}$ to fit $\overline{\code{D}}_{sj}$ through a $k$-fold cross-validation learning pipeline. The resulting learners $l^{*}$ (i.e., the fitted regression models) are stored into a set $O$ together with other machine learning output material. When having a number of learner models available in $O$, the procedure \codeSmall{ {data\_correction}} takes the best one $l^*$ for a target sensor $s$ in order to correct a given dataset $\overline{\code{D}}_{sk}$ of a $k$ observation period from the LCAWS's database $\overline{\code{DB}}_{L}$. The resulting dataset $\overline{\code{D}}_{sk}^{~*}$ containing the corrected weather parameters is thus stored in a final database  $\overline{\code{DB}}_{L}^{~*}$.}

{The proposed intelligent sensor calibration and data correction is expected to be a continuous process. This means that, once we have a set of fitted models in $O$, the regression models can be applied to correct data as soon as new weather parameters are available in $\overline{\code{DB}}_{L}$. Meanwhile, the set $O$ should not be a permanent repository of regression models, so that the procedure \codeSmall{ {learning\_pipeline}} has to be run whenever new ground-truth values are available in $\overline{\code{DB}}_{P}$. In Section~\ref{sec:LCAWS-calibration}, we describe details of the implementation and methodology the proposed calibration and data corrections, as well as the discussion of results we obtained from experiments conducted exhaustively.}

\section{Weather Sensor Data Processing}
\label{sec:LCAWS-dataprocessing}

In this section, we describe the equations we implemented throughout the procedure \codeSmall{  {summary()}} to process the different sensor data and obtain the desired weather parameters.

\subsection{Digital Sensor Measurements}

Atmospheric pressure~(AP), air temperature~(AT), and relative humidity~(RH) are measured by the Bosch BME280 digital sensor kept in the sensor housing. The BME280 allows for accomplishing a period of measurement with configurable oversampling. After such a measurement, an infinite impulse response filter (IIR) can be applied to AP and AT values in order to increase the output signal resolution from 16 to \unit[20]{bits} while reducing bandwidth and removing short-term fluctuations. RH measurement does not fluctuate rapidly, thus it requires no filtering. BME280s IIR is given by a   {low-pass filter:}
\begin{equation}
    x_i^* = \frac{x_{i-1}^* \cdot (2^k - 1) + x_i} {2^k},
\label{eq:f_f}
\end{equation}
where $x_i$ is the observed measurement from the ADC output data; $x_{i-1}^*$ is the previous measurement filtered; and $2^k$ is a filter coefficient with a factor $k = [0, 4]$. 

The LCAWS oversampling setting is of {16$\times$} 
measurements with filter disabled (i.e., $k=0$) per data reading. A cycle of a single sampling takes the time of three consecutive measurements (AT, AP, and RH) followed by a standby time of \unit[$0.5$]{ms}. The LCAWS client data collector accomplishes one data reading per minute from the BME280 module. Thus, the mean of filtered measurements per hour for a BME280 sensor is:
\begin{equation}
\overline{\text{X}} = \frac{1}{N} \sum_{i=1}^{N} x_i^*,
\label{eq:mean_BME280}
\end{equation}
where $N$ is the sample size of measurements obtained in a period of 1 h observation; and $x_i^*$ is an $i$ measurement filtered with Equation~(\ref{eq:f_f}).

\subsection{Rainfall Precipitation}

The LCAWS pluviometer is a tipping bucket rain gauge, thus it counts cumulatively the number of magnetic pulses whenever the bucket tips due to the rainfall precipitation. From the number of pulses, the correspondent rainfall precipitations in millimeters   {per hour is:}
\begin{equation}
\overline{\text{RG}} = \sum_{i=1}^{N} f_{\text{d}}(p_{i}) \times \nu,
\label{eq:mean_P}
\end{equation}
where $N$ is  the  sample  size  of  measurements obtained in 1 h observation; $p_i$ is the number of pulse clicks accumulated in that measurement; $\nu$ is the amount of precipitation measured by each pulse generated by the tipping of the LCAWS bucket, which is \unit[$\nu=\text{0.25}$]{mm}; and $f_{\text{d}}(x)$ is the lagged differences between two consecutive values, $i$ and $i+1$ in~$x$. 

The lagged differences are given by:
 \begin{equation}
 f_{\text{d}}(x_i) = 
\begin{cases}
    x_{i+1} +  x_\text{max} - x_{i}, & \text{if } (x_{i+1}  - x_{i}) < 0, \\
    x_{i+1}  - x_{i},  & \text{otherwise},
\end{cases}
\label{eq:d}
 \end{equation}
where $x_\text{max}$ is the maximum unsigned integer value in which $x_i$ can store. The LCAWS client data collector module assumes 16-bit integer on the Arduino Mega microcontroller, i.e., $x_\text{max} = 2^{16} - 1$.

\subsection{Wind Speed}

The raw measurement obtained from the LCAWS anemometer computes cumulatively the number of revolutions of its cup-mounted arm. To convert such a data into wind speed in meters per second, $w$, the data processing is accomplished   {as follows:}
\begin{equation}
w =  C  \cdot R,
\label{eq:WS_lc}
\end{equation}
where $C = \pi \times o$ is the circumference of a revolution for a propeller of $o$ diameter; and $R$ is the expected amount of revolutions of the propeller observed in a second. The LCAWS anemometer provides revolutions of circumference \unit[$C=\text{92.4}$]{cm}, once its arms have a diameter of \unit[$o=\text{29.4}$]{cm}. 

Since a cup-mounted arm revolution causes a magnetic pulse, $R$ provides the expected number of pulses per second,   {which is given by:}
 \begin{equation}
R =  \frac{f_{\text{d}}(r)}{f_{\text{d}}(t)} \cdot s,
 \end{equation}
where $f_{\text{d}}(x)$ is given by Equation~(\ref{eq:d}); $r$ the set of cumulative pulses observed in a 1 h period; $t$ is a set of the numbers of cumulative milliseconds passed since LCAWS datalogger started running; and $s$ is the time unit in seconds, i.e., \unit[$s = \text{1000}$]{ms}. 

Cup anemometers and wind vanes are typical weather instruments that provide horizontal wind measurements, so that wind speed is related to a particular wind direction. Thus, to obtain the mean of wind speed of an observation period, the east--west and north--south wind directions require to be mapped into the Cartesian plane of two components, $x$ and $y$. The mean of the vector components, $\bar{x}$ and $\bar{y}$,   {are given by:}
\begin{equation}
\bar{x}  = \frac{1}{N}  \sum_{i}^{N} \left[ -w_i \cdot \sin \left( 2 \pi \cdot \frac{\theta_i}{360} \right) \right],
\label{eq:bar_x}
 \end{equation}
 \begin{equation}
\bar{y}  = \frac{1}{N}  \sum_{i}^{N} \left[ -w_i \cdot \cos \left( 2 \pi \cdot \frac{\theta_i}{360} \right) \right],
\label{eq:bar_y}
\end{equation}
where $N$ is the sample size of measurements obtained with the weather station in 1-hour observation; $w_i$ is the wind speed in m/s; and $\theta_i$ is the wind direction in degrees. The wind speed $w_i$ and LCAWS direction $\theta_i$ are given by Equations~(\ref{eq:WS_lc}) and~(\ref{eq:f_theta}), respectively.

From the means of vector components $x$ and $y$, the mean of wind speed is determined from the resultant vector average:
\begin{equation}
\overline{\text{WS}}  = \sqrt{\bar{x}^2 + \bar{y}^2},
\label{eq:mean_WS}
\end{equation}
where $\bar{x}$ and $\bar{y}$ are the component averages obtained with Equations~(\ref{eq:bar_x}) and~(\ref{eq:bar_y}), respectively.

\subsection{Wind Vane}

The LCAWS wind vane operates through eight reed switches fixed to the chassis, which are switched on by a permanent magnet fixed to the rotary part of the vane. This gives the wind vane a resolution of $\theta_{\text{res}} = \ang{45}$, corresponding to the Cardinal directions north, northeast, east, southeast, south, southwest, west, and northwest.

In the magnetic sensor device, the wind direction corresponds to the current voltage of the vane voltage divider in the continuous interval of 0{$v$} and 5{$v$}. ADC provides such a voltage in 10-bit integer values, which can be lossless reduced to the space of an 8-bit integer. Thus, the value domain is within ${V}=[0, {v}_\text{max}]$, where ${v}_\text{max} = 2^8 - 1$. Then, a voltage ${v} \in {V}$ is mapped into one out of nine discrete vane positions ${p} \in {P}$. A mapping function $f_{\text{p}}: {V} \rightarrow {P}$ is   {given by:}
\begin{equation}
f_{\text{p}}(\text{v}) = \biggl\lfloor R_\text{ref} \cdot \left(\frac{\text{v}_\text{max}}{\text{v}} -1 \right) \biggl\rceil,
\label{eq:f_pos}
\end{equation}
where \unit[$R_\text{ref} = 4700$]{$\Omega$} is the reference resistor.

If $1 \leq {p} \leq 8$, then the vane position $p$ can be mapped into a corresponding angle $\theta \in \Theta$, where $\Theta = \{0,  45, \cdots, 315 \}$ is a set of angles spaced in $\theta_{\text{res}}$ degrees according to the eight expected Cardinal directions. A mapping function, $f_\theta: {P} \rightarrow \Theta$,   {is given by:}
\begin{equation}
f_\theta ({p}) = 
\begin{cases}
    \theta_0,	& \text{if } {p} = 0, \\
    \theta_\text{res} \cdot ({p} - 1),		& \text{otherwise},
\end{cases}
\label{eq:f_theta}
 \end{equation}
where $\theta_0 = \ang{225}$ is the calibration of a particular position of the vane when $f_{\text{p}}({v}) = 0$, in which the circuit is spliced so that ACD cannot provide a valid voltage; and $\theta_\text{res} = \ang{45}$ is the angular space between two intercardinal directions.

When determining $\theta$, the mean of wind direction, $\overline{\text{WS}}$, of a period of observation  { is determined by:}
 \begin{equation}
\overline{\text{WD}} = \arctan \left( \frac{\bar{x}}{\bar{y}} \right) \cdot \frac{180}{\pi} + 180,
\label{eq:mean_WD}
\end{equation}
where $\bar{x}$ and $\bar{y}$ are the means of the vector wind components obtained from Equations~(\ref{eq:bar_x}) and~(\ref{eq:bar_y}), respectively.

\section{Weather Station Validation}
\label{sec:LCAWS-validation}

To validate LCAWS, we applied an experimental methodology of three main steps: {weather station deployment}, {data acquisition}, and {data analysis}. In the next subsections, we discuss each step. 

\subsection{Weather Station Deployment}

\begin{figure*}[]
    \centering
    \subfigure[~Deployment location (red point): 23$^\text{o}$09'21.5'' S 45$^\text{o}$7'30.0'' W.]{  
        \includegraphics[scale=0.52, trim = {2cm 16.7cm 2cm 1.5cm}]{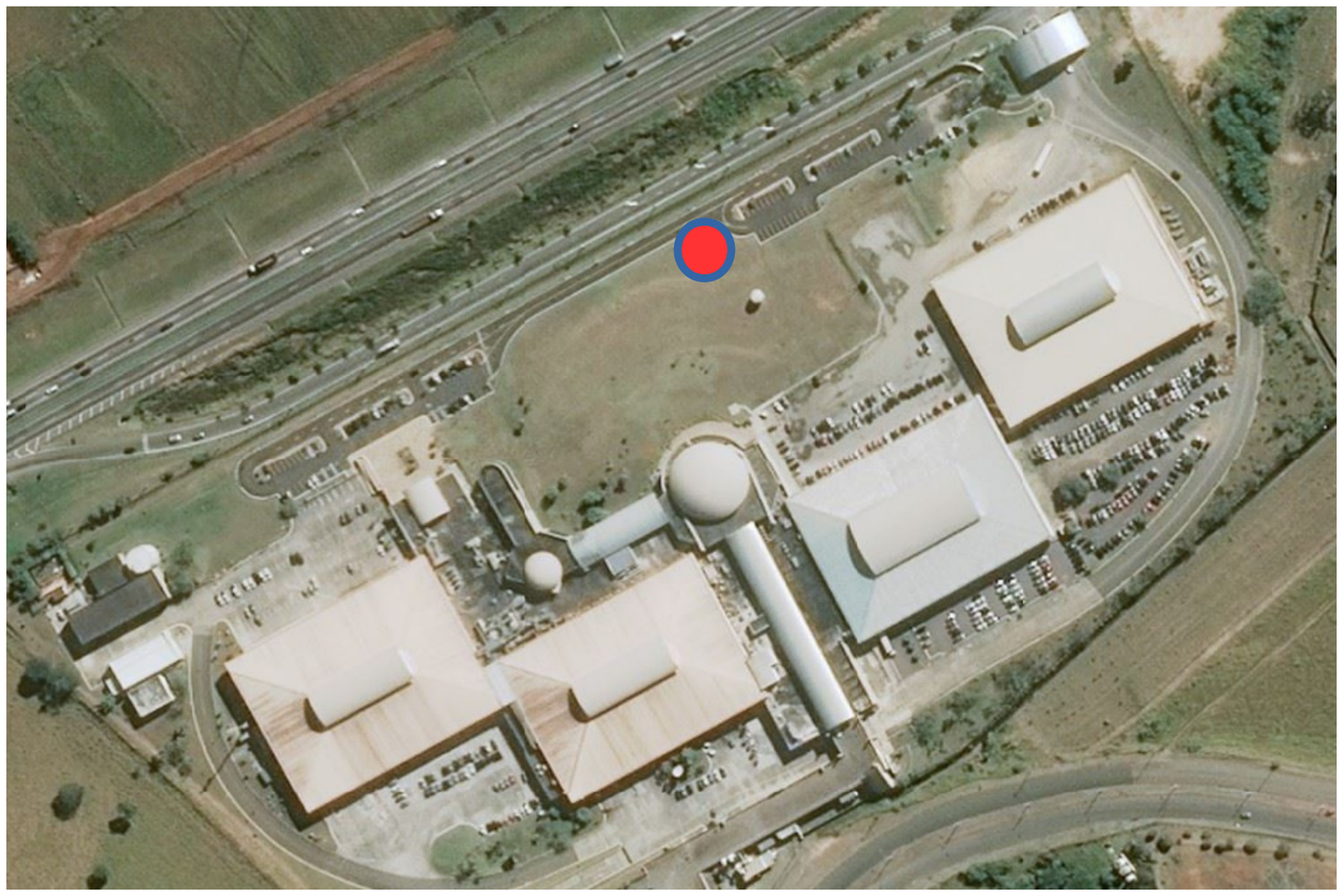}
    } 
    \hspace{-0.2cm}
    \subfigure[~LCAWS.]{
        \includegraphics[scale=0.0395]{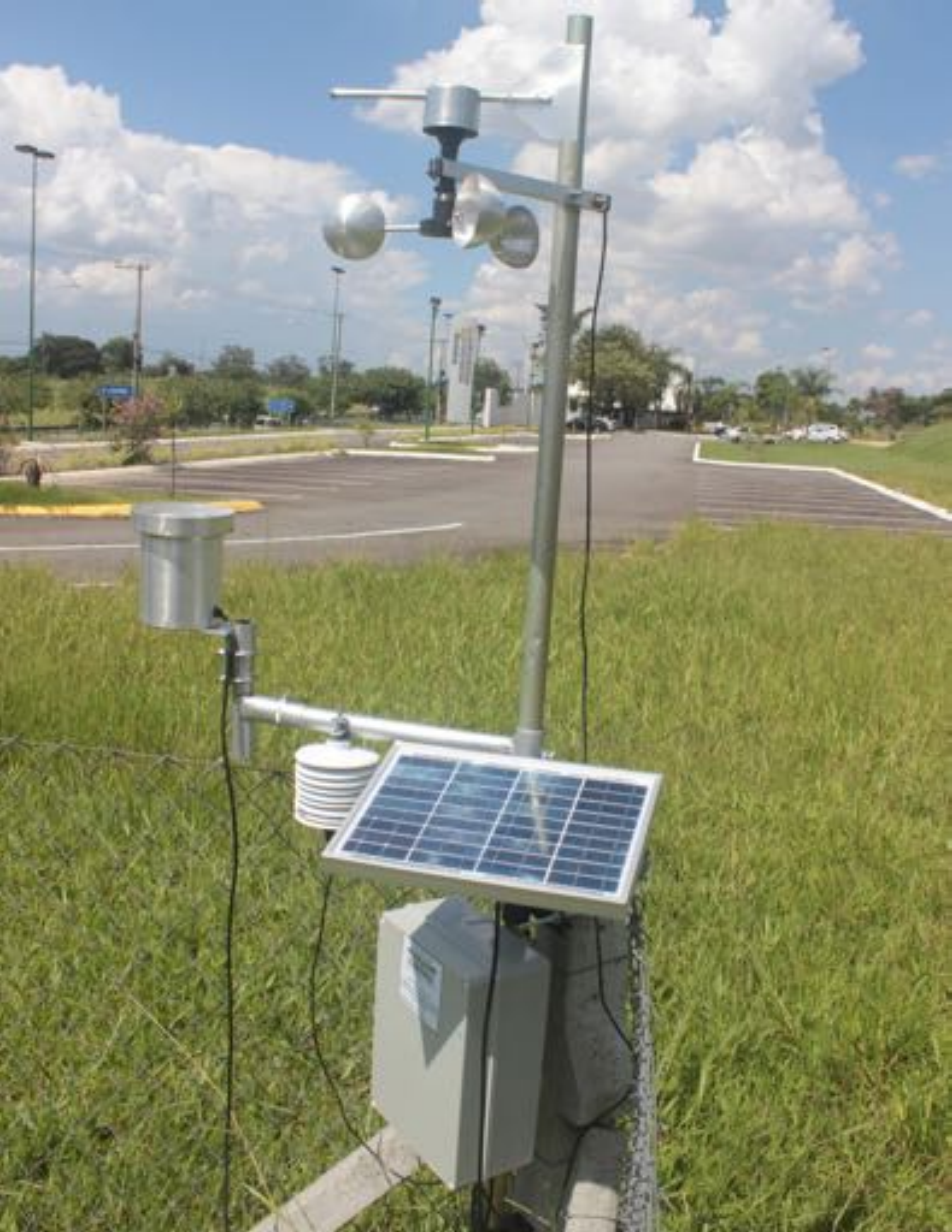}
    }
    \hspace{-0.2cm}
    \subfigure[~PWS.]{
        \includegraphics[scale=0.0582]{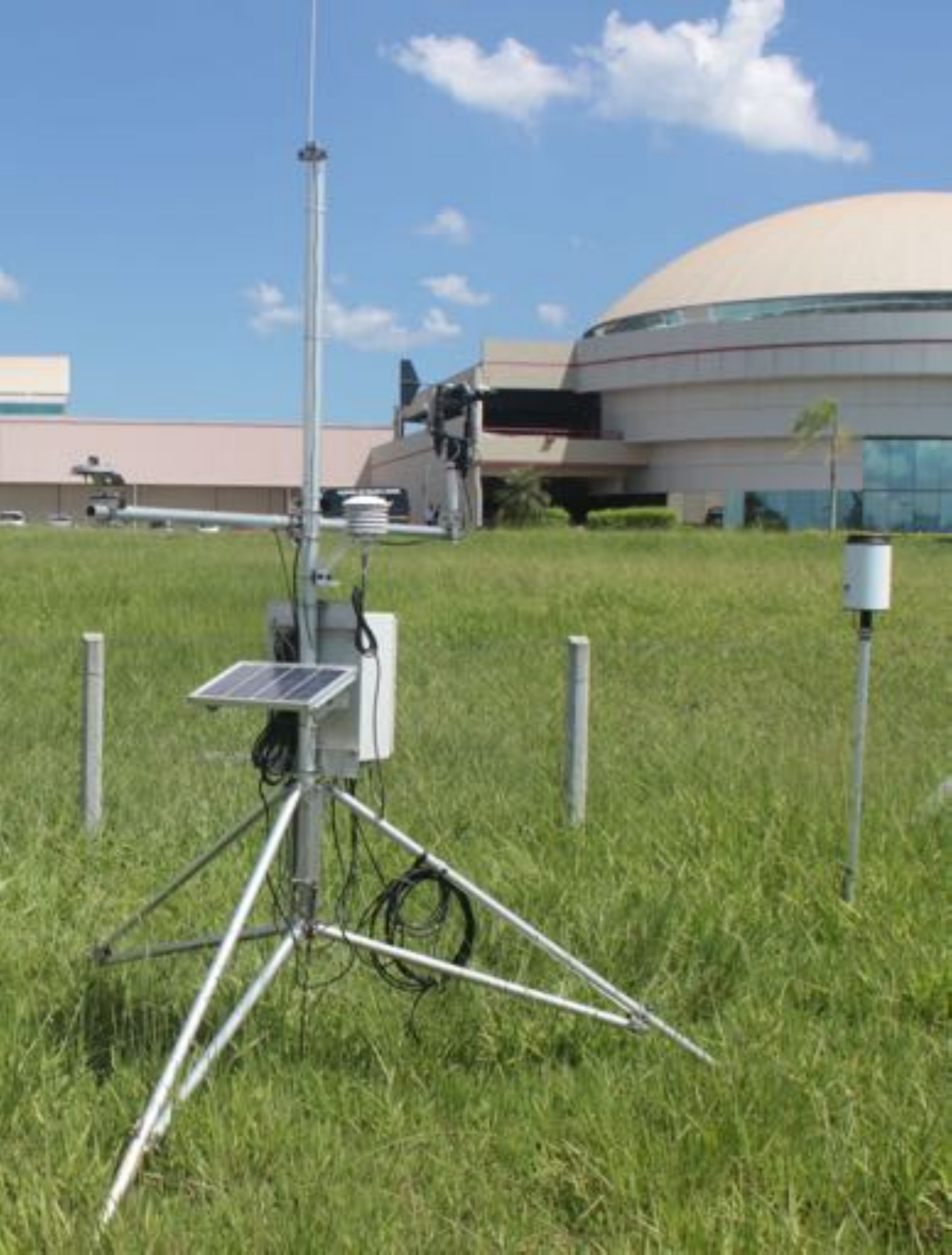}
    }
    \caption{Weather station deployment {in situ} pictures: ({a}) a \unit[300]{ft} aerial image of the São José dos Campos Technological Park (PqTec)\cite{pqTec:2020} -- image adapted from OpenStreetMap {contributors}~\cite{Deployment.Location:2020}, with the red point marking the exact location (-23.15598 -45.79166) of the deployment; ({b}) the Low-Cost Automatic Weather Station (LCAWS); and ({c}) the Professional Weather Station (PWS).}
    \label{fig:deployment-photos}
\end{figure*}

We deployed LCAWS at the outdoor test environment of the Technology Park~(PqTec)~\cite{pqTec:2020, Deployment.Location:2020} of the city of São José dos Campos, state of São Paulo, southeast Brazil. In order to asses the weather measurement results, LCAWS was co-located 3 m apart of a reference weather station. Both stations were able to catch and measure the same weather events at near location and time, allowing direct comparison of the measurement results between them. The 
Professional Weather Station~(PWS) of reference we utilized was the Campbell Scientific   a{CR200 Series}  {(Campbell Scientific, Inc, Leicestershire,
UK)}~\cite{CR200:2015}. Specifications of both stations are detailed in \mbox{Table~\ref{tab:both-specifications}}. Figure~\ref{fig:deployment-photos} shows the pictures of the deployment location.

\begin{table}[h]
\centering
\footnotesize
\caption{Weather stations' specifications.}
\label{tab:both-specifications}
\setlength{\tabcolsep}{1pt}
\begin{tabular*}{\linewidth}{ @{\extracolsep{\fill}} p{2cm} p{3.6cm} p{3cm}}
\toprule
                & \textbf{LCAWS} 
                & \textbf{PWS} \\
\cmidrule{2-3}
Manufacture:    & Multi-suppliers 
                & Campbell Scientific \\
Model:          & Beta v3.0 
                & CR200(X) \\
Maturity level: & Academic/test prototype 
                & Professional use \\
Weather sensors:& AP, AT, RH, RG, WS, WD 
                & AP, AT, RH, RG, WS, WD \\
Control sensors: & GPS, BME280, Battery l.
                & Battery level \\
API languages:  & C++, C, R, Python 
                & CRBASIC \\
A/D converter:  & 10 bits 
                & 12 bits \\
Max. scan rate: & 1/s
                & 1/s \\
User program:   & -- 
                & PC200W software \\
Processing:     & Arduino Mega 2560 
                & CR200(X) CPU \\
Memory:         & \unit[8]{KB} RAM,
                & \unit[6.5]{KB} program storage \\
                & \unit[256]{KB} program storage 
                & \\
Data storage:   & \unit[2]{GB} microSD 
                & \unit[128-512]{KB} EEPROM \\
Data format:    & C++ primitive data types
                & \unit[4]{B} per data point \\
Data retrieval: & USB/RS232, GPRS, microSD
                & RS232, PCCOM \\
Data correction: & ML-based sensor calibration
                & -- \\
Communication:  & USB/RS232, GPRS shield 
                & Serial RS232 \\
Temp. range:    & -- 
                & \unit[-40$^\text{o}$]{C} to \unit[+50$^\text{o}$]{C} \\
Datalogger      & Plastic, IP55,
                & Aluminium, NEMA 4X, \\
enclosure:      & \unit[30]{cm}$\times$~\unit[22]{cm}~$\times$~\unit[12]{cm}
                & \unit[14]{cm}~$\times$~\unit[17.6]{cm}~$\times$~\unit[5.1]{cm} \\
Current drain:  & \unit[80]{mA} avg, \unit[2]{A} peak (GPRS TX), \unit[12]{V}
                & \unit[3]{mA} avg, \unit[75]{mA} peak, \unit[12]{V} \\
Battery:        & \unit[6]{Ah}
                & \unit[0.08]{Ah} \\
Power supply:   & Solar panel \unit[55]{Wp} 
                & Solar panel \unit[10]{Wp} \\
Redundancy:     & --
                & Battery backup \\
Protections:    & Moisture monitoring
                & EMI, ESD \\
\bottomrule
\end{tabular*}
\end{table}

\subsection{Weather Data Acquisition}

\begin{figure*}[h]
\centering
	    \includegraphics[scale=0.73, trim = 0.2cm 0cm 0cm 0cm]{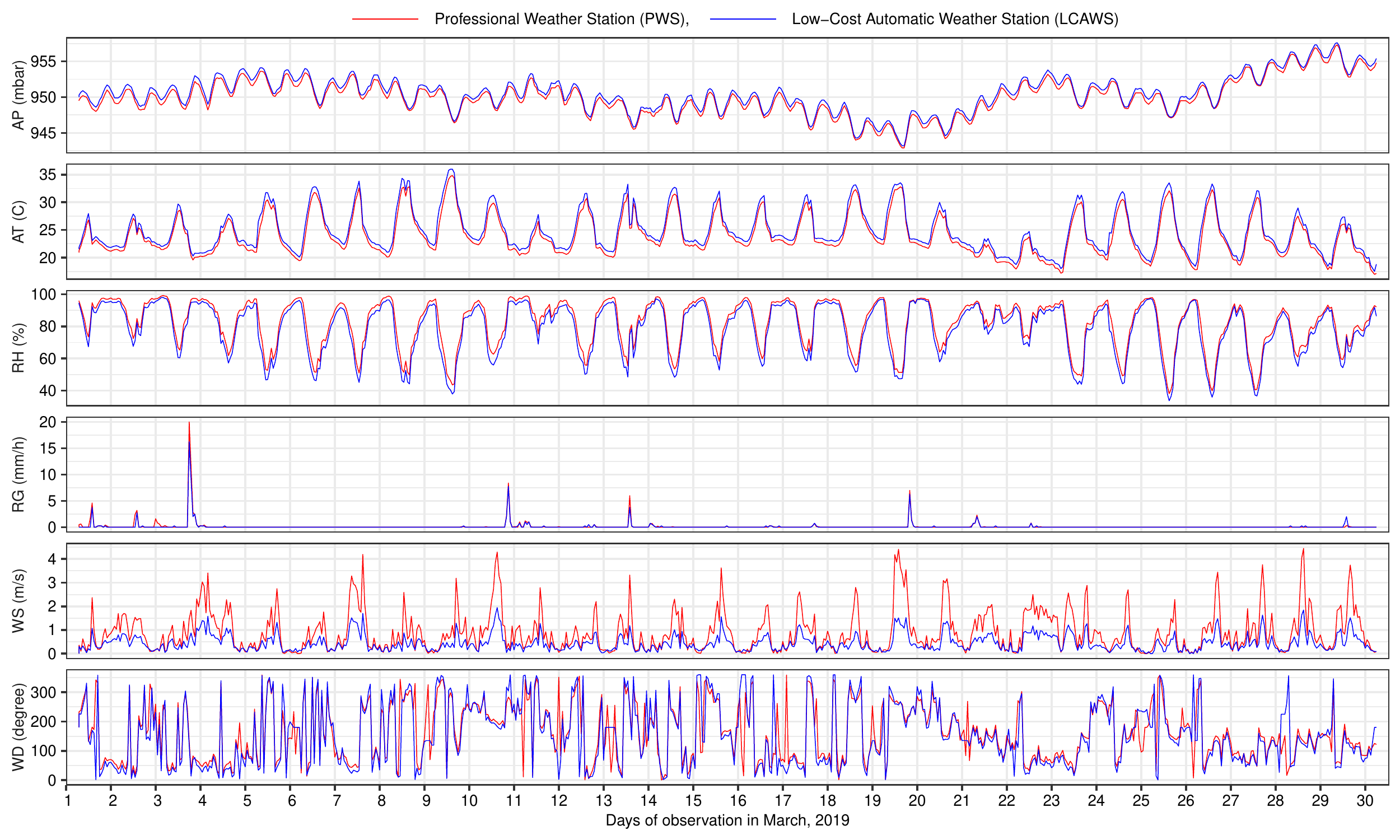}
\caption{Time-series of 30 days of weather observation with both stations co-located side-by-side, the Low-Cost Automatic Weather Station (LCAWS) and the Professional Weather Station (PWS), from different sensors: Air Pressure~(AP), Air Temperature~(AT), Relative Humidity~(RH), Rain Gauge~(RG), Wind Speed~(WS), and Wind Direction~(WS).}
\label{fig:raw-time-series}
\end{figure*}

We collected the weather measurements from both stations during 30~days of continuous observations, during the rainy season in southeast Brazil (March 2019). Each station provided the measurements once per minute, so that we processed the corresponding weather parameters into statistics per hour, according to the data processing described in Section~\ref{sec:LCAWS-dataprocessing}. Thus, from a large uninterrupted observation period, we obtained samplings of atmospheric pressure (AP), air temperature (AT), relative humidity (RH), rain gauge (RG), wind speed (WS), and wind direction (WD). Figure~\ref{fig:raw-time-series} shows the time-series for each parameter provided by LCAWS and PWS during the 30~days of observations. When interposing the results of the stations, the time-series gives us a description of how each parameter had behaved and how close the results of LCAWS are with respect to PWS. Since we used PWS as a reference, we assume its results as being the ground-truth to discuss closeness and dissimilarity from the results produced by LCAWS. 

{For each sensor of each station, we generate a dataset by summarizing statistics (e.g., minimum, 1st quartile, median, mean, 3rd quartile, maximum, sum, and mode) per hour over the corresponding weather parameter (i.e., over the weather data processed with the equations described in Section~\ref{sec:LCAWS-dataprocessing}). Thus, the summarized weather parameter is described in the domain of non-negative real numbers, where each tuple in the dataset is driven to the corresponding timestamp $t_\code{ts}$. The datasets are publicly available
and can be found in~\cite{complementary.material:2021}.}

\subsection{Weather Data Analysis}

{We analyzed the parameters by computing various performance metrics such as Pearson's Correlation Coefficient~(PCC), R-squared coefficient of determination~(R2), and residuals analysis from Mean Squared Error~(MSE) and Root Mean Squared Error~(RMSE). In addition, we applied $t$-tests over the obtained results in order to verify the statistical~significance.}

To correlate the weather parameters in pairs of sensors, 
 {we apply PPC:}
\begin{equation}
\text{PPC} = \frac{\sum_{i=1}^N (x_i - \bar{x}) \cdot (y_i - \bar{y})}{\sqrt{\sum_{i=1}^N (x_i - \bar{x})^2} \cdot \sqrt{\sum_{i=1}^N (y_i - \bar{y})^2}},
\end{equation}
where $x$ and $y$ are two given values correlated from two sensors, and $\bar{x}$ and $\bar{y}$ are their averages, respectively. 

Considering the PWS results as being the ground-truth, we determine the goodness-of-fit from LCAWS to PWS for each pair of equivalent sensors by applying the R2 coefficient, 
\begin{equation}
    \text{R2} = 1 - \frac{\sum_{i=1}^{N}{(y_i - \hat{y}_i)^2}}{\sum_{i=1}^{N}{(\bar{y} - y_i})^2}, 
    \label{eq:R2}
\end{equation}
where $y_i$ and $\hat{y}_i$ are the actual and predicted values, i.e., PWS and LCAWS results, respectively. The best case is $\text{R2} = 1$, i.e., when the predicted values exactly match the actual ones, so that the residual sum of squares is zero. 
 
To measure how well the results of LCAWS can fit to PWS, for each pair of equivalent sensors, we analyze the residuals between them from the average of squared errors and square root of such an average, respectively,
\begin{eqnarray}
    \text{MSE} &=&\frac{1}{N} \sum_{i=1}^{N}{(y_i - \hat{y}_i)^2}, \\
    \text{RMSE} &=&\sqrt{\text{MSE}},
\end{eqnarray}
where $y_i$ and $\hat{y}_i$ are the actual and predicted values, as in Equation~\eqref{eq:R2}. While MSE is a risk metric corresponding to a quadratic score from the average residual magnitude, RMSE is its squared root, which can be interpreted with the same unit as the measured data. 

\begin{figure}[h]
\centering
	\hspace{-0.3cm}
    \subfigure[][LCAWS]{
			\includegraphics[scale=0.328, trim = {0.9cm 0.2cm 0.5cm 0cm}, clip]{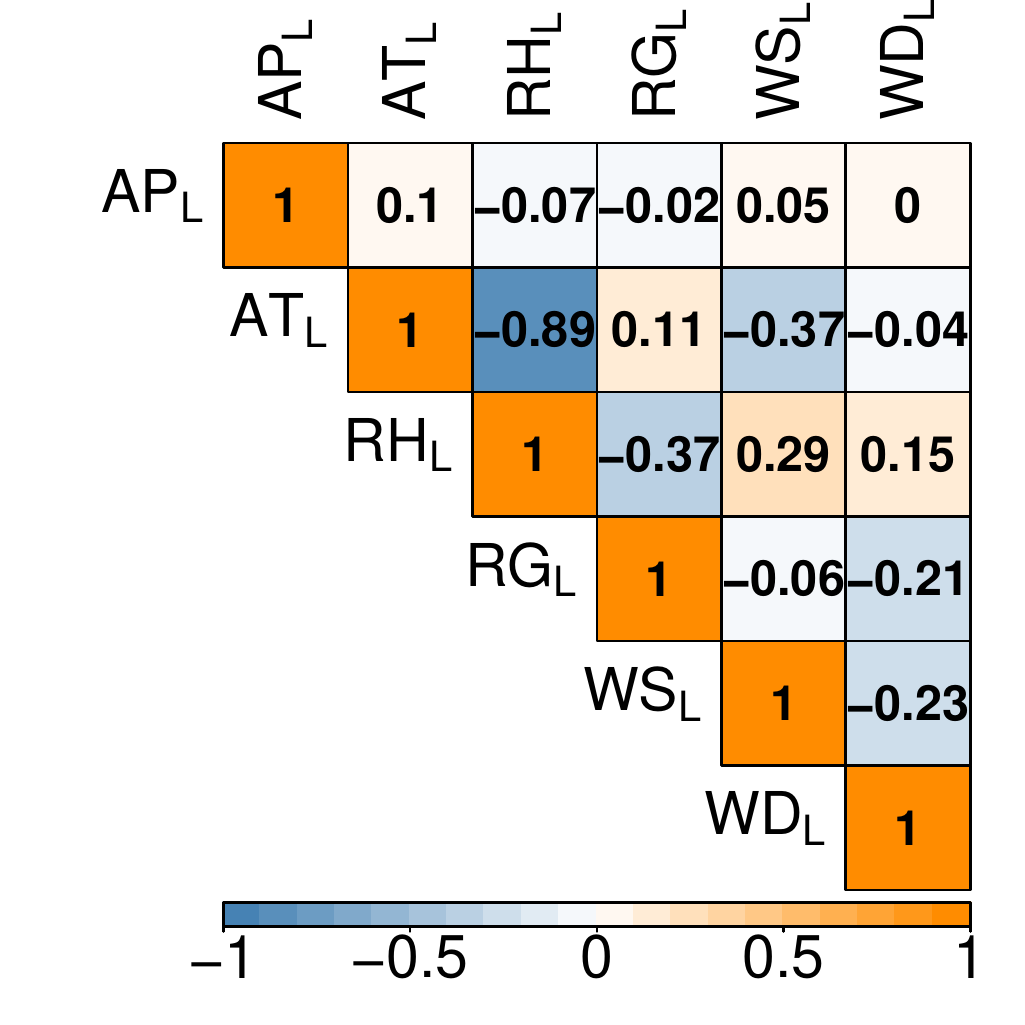}
		}
	\hspace{-0.34cm}
	\subfigure[][PWS]{
			\includegraphics[scale=0.328, trim = {0.9cm 0.2cm 0.5cm 0cm}, clip]{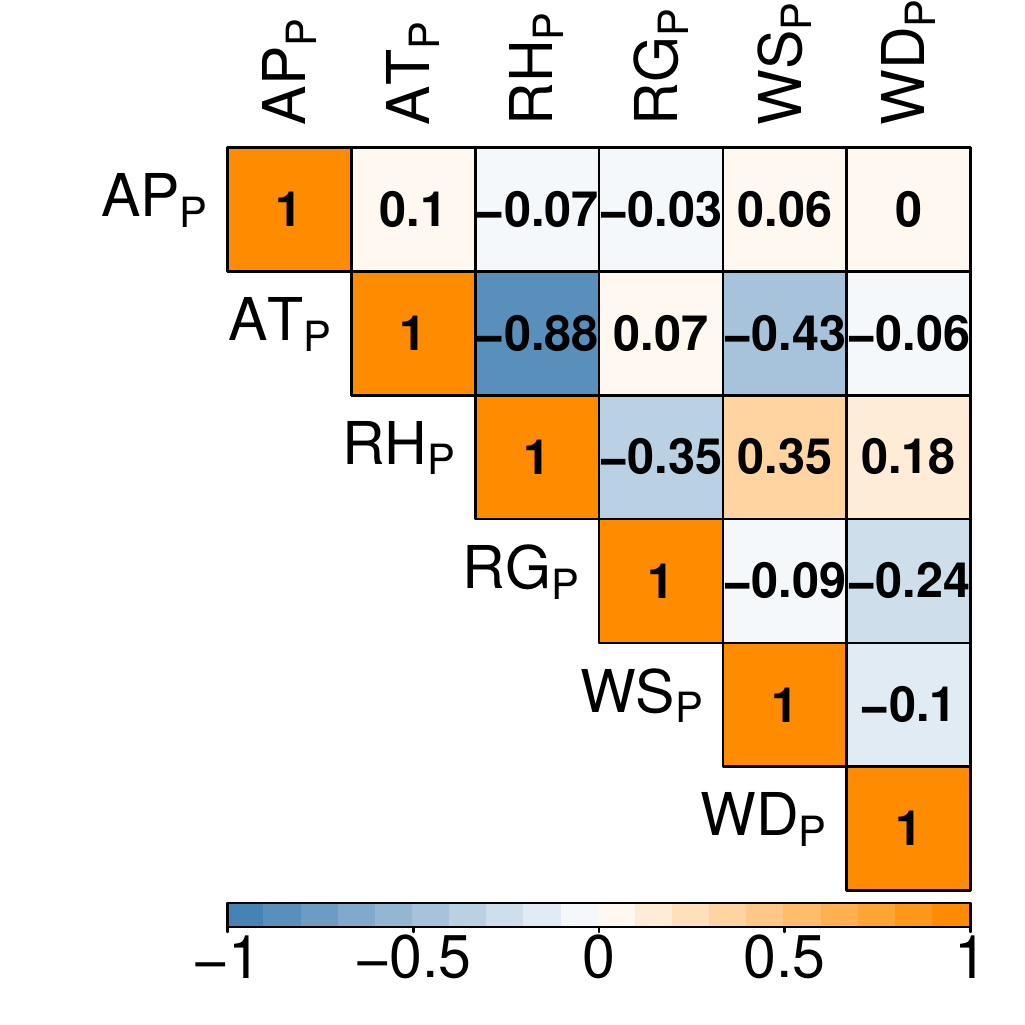}
		}
	\hspace{-0.34cm}
	\subfigure[][LCAWS vs. PWS]{
			\includegraphics[scale=0.328, trim = {0.9cm 0.2cm 0.5cm 0cm}, clip]{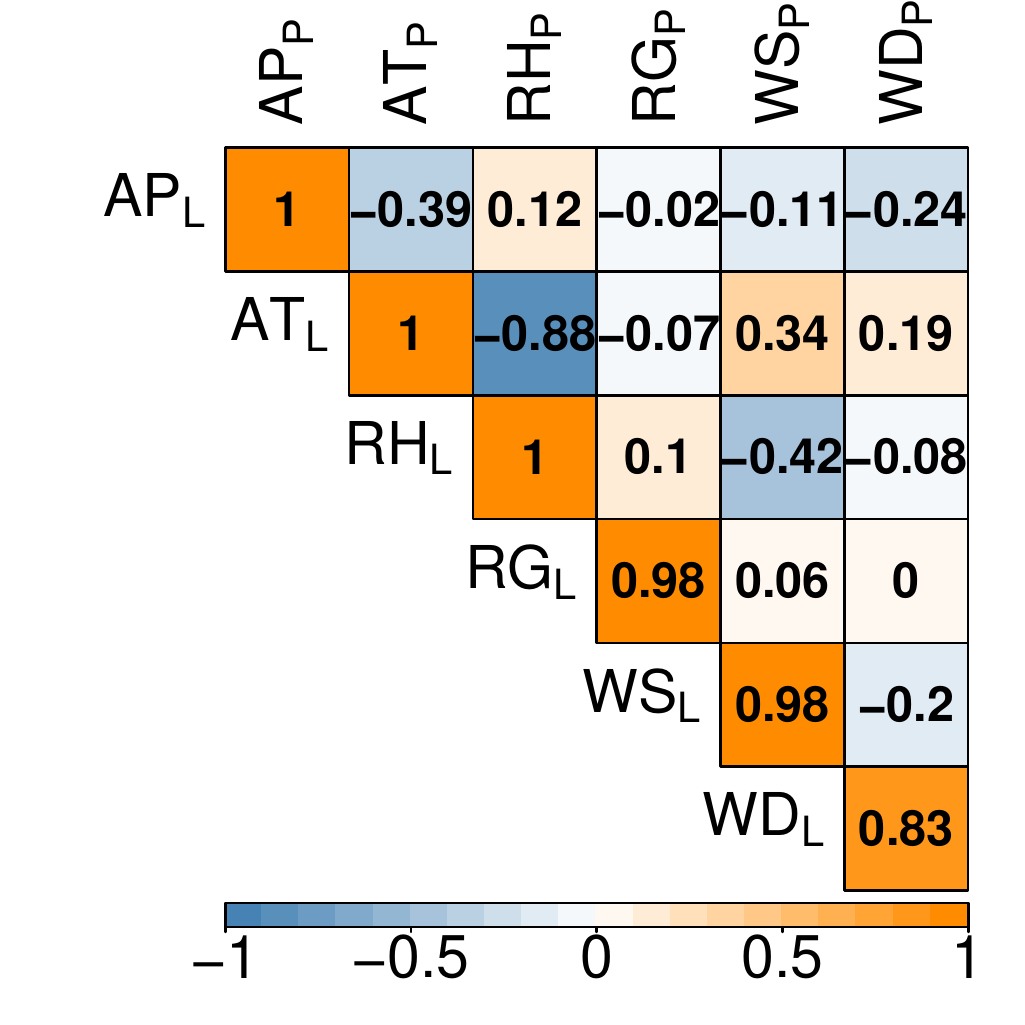}
		}
\caption{Pearson correlation coefficient (PCC) of the observed results (30 days): ({a}) between sensors of LCAWS, ({b}) between sensors of PWS, and ({c})~between sensors of LCAWS and PWS.}
\label{fig:raw-pearson-cor}
\end{figure}

Figure~\ref{fig:raw-pearson-cor} shows the PPC results between pairs of sensors. From the PPC similarity of internal sensor pairs in each station, as shown in Figure~\ref{fig:raw-pearson-cor}a,b, one can assume that the results between the stations are coherent. While there was no correlation between AP and the other sensors, RH was strongly and negatively correlated to AT. The other pairs of sensors had positive and negative PPC results, however, with no strong coefficients greater than 0.5 or lower than $-$0.5. In general, the obtained PPCs show that LCAWS works in accordance with PWS. However, when combining sensors in pairs between LCAWS and PWD, as shown in Figure~\ref{fig:raw-pearson-cor}c, we observed correlation divergences. While the Bosch BME280s digital sensors (AP, AT, RH) allowed coefficients closer to the ones in PWD, the magnetic-based sensors (RG, WS, WD) could not provide such stronger correlations.

To investigate such a divergence, we compared those pairs of sensors by analyzing the residual results between them with the performance metrics R2, MSE, and RMSE. Table~\ref{tab:raw-stat} presents the metrics obtained between LCAWS and PWS. To assess the statistical significance of the results, we applied paired $t$-tests by assuming the following parameterization: the confidence interval of \unit[95]{\%}, with the null hypothesis $\text{H}_0 : s_\text{L} = s_\text{P}$, i.e., a datum collected from a given sensor $s$ in LCAWS is equal to a corresponding one in PWS. Otherwise, the alternative hypothesis $\text{H}_1: s_\text{L} \neq s_\text{P}$.

\begin{table}[h]
\footnotesize
\centering
\caption{Performance metrics and statistical significance between LCAWS and PWS (30-day observation).} 
\label{tab:raw-stat}

\begin{tabular}{llllrl}

  \toprule \textbf{Sensor} & \textbf{R2} & \textbf{MSE}  & \textbf{RMSE} & \multicolumn{2}{c}{\textbf{Significance}} \\ & & & & {\textit{\textbf{t}}\textbf{-value}} & {\textit{\textbf{p}}\textbf{-value}} \\ \midrule
  
AP & 0.9557 & 0.2815 & 0.5305 & 3.7500 & ** \\ 
  AT & 0.9260 & 0.9789 & 0.9894 & 4.6400 & ** \\ 
  RH & 0.9186 & 17.3133 & 4.1609 & $-$4.5700 & ** \\ 
  RG & 0.9390 & 0.0660 & 0.2569 & $-$0.5100 & 0.6132 \\ 
  WS & 0.3445 & 0.4979 & 0.7056 & $-$13.0100 & ** \\ 
  WD & 0.6136 & 3567.6384 & 59.7297 & $-$1.1300 & 0.2601 \\ 
   \toprule \multicolumn{6}{l}{Significance code: (**) \textit{p}-value $\leq$ 0.001.}
\end{tabular}
\end{table}

\begin{figure*}[]
\centering
	     \includegraphics[scale=0.73, trim = 0.2cm 0cm 0cm 0cm]{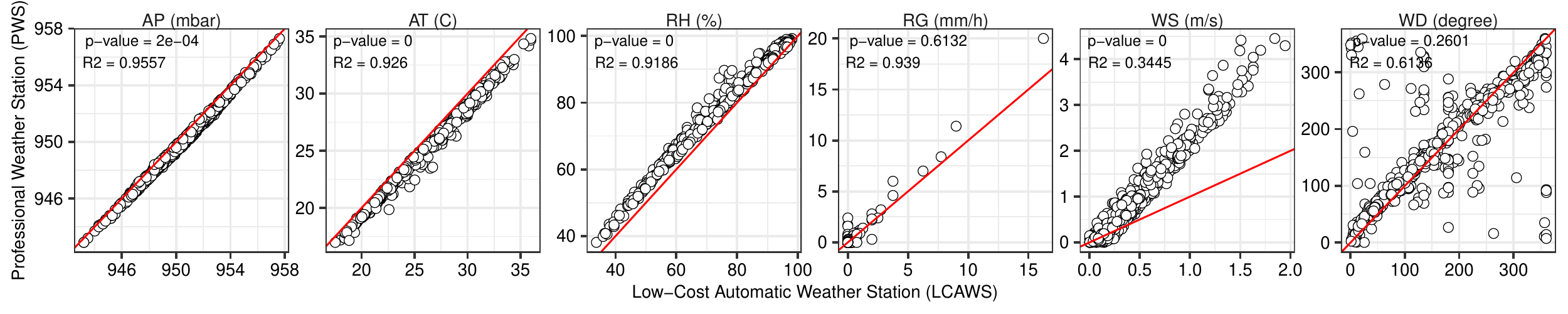}
\caption{Scatter plots of 30-day observation period with the Low-Cost Automatic Weather Station (LCAWS) and the Professional Weather Station (PWS) from different sensors: Air Pressure~(AP), Air Temperature~(AT), Relative Humidity~(RH), Rain Gauge~(RG), Wind Speed~(WS), and Wind Direction~(WS).}
\label{fig:raw-scatter-plots}
\end{figure*}

As presented in Table~\ref{tab:raw-stat}, the combined digital sensors in Bosch's BME280 module could allow high accuracy, with $\text{R2} \geq 0.91$ and small residuals according to the obtained MSE and RMSE. This can be seen in the scatter plots in Figure~\ref{fig:raw-scatter-plots}, where the results for AP, AT, and RH from LCAWS were close to the ones obtained with PWS. Although obtaining promising performance metrics, when assessing the significance of the results between LCAWS and PWD, we verified \textit{p}-values of high significance, lower than 0.001 (**). In other words, the null hypothesis is rejected for the sensors AP, AT, and RH, which means that statistically such sensors provide different results. In case of accepting H$_0$ with a tight match between the outcomes of pairs of sensors, we would have $\textit{t}\text{-value} \approx 0$ and $\textit{p}\text{-value} \approx 1$.

Different from the digital sensors, the rain gauge (RG), wind speed (WS), and wind direction  {(WD)} 
 are magnetic sensors whose accuracy depends on the efficiency of analog apparatus. Weather events are measured from reed switch pulses caught from the mechanical interaction caused by those events on the sensor device. In addition, these sensors are implemented under different designs for each station, which include different physical dimensions. The amount precipitation needed to tip the LCAWS tipping bucket is higher than that of the PWS (\unit[0.25]{mm} vs. \unit[0.10]{mm}). This means that the PWS rain gauge sensor is more sensitive, by requiring less precipitation to catch rainy events than the LCAWS sensor. This implied different accounting of precipitation by the stations during the 30-day observation period, with \unit[80.75]{mm} and \unit[98.7]{mm} registered by LCAWS and PWS, respectively. Although the PWS bucket resolution was about 2.5 times higher, the rain gauge residuals between the stations were of low magnitude, with MSE~=~0.066 and RMSE~=~0.2569 (mm), while allowing predictability of R2 $\geq$ 0.93, as described in Table~\ref{tab:raw-stat}. Such an RMSE represents an average error of $\pm$1 LCAWS bucket between the stations. From these residuals, the \textit{p}-value was of 0.61 so that there is no significant difference between the rain gauge measurements of
PWD and LCAWS. However, it is important to highlight that the rain gauge time-series are described by 91.4{\%} and 88.8{\%} of non-rainy hours for LCAWS and PWS, respectively. This unbalanced sampling of rainy and non-rainy measurements also explains the low residuals, even when sensors have different bucket volumes.

Among the magnetic sensors, the worst performance metrics came from the wind speed (WS) sensor, with the lowest R2 of 0.34, and residuals of $\text{MSE} = 0.50$ and \unit[$\text{RMSE} = 0.70$]{m/s}. Regarding the space of the observed results in the range \unit[$0 \geq \text{WS} \leq 4.45$]{m/s}, those residuals are quite representative. However, from the time-series in Figure~\ref{fig:raw-time-series}, we observe that the faster the wind speed, the higher the residuals between the WS results. On the other hand, while the residuals are of different scales, they behave under a similar pattern with correlation $\text{PPC} = 0.98$.

When comparing the results of the wind direction sensors (WD) between the stations, a \textit{p}-value~greater than 0.05 accepts the null hypothesis. However, it lies on the higher magnitude of residuals with MSE~=~3567.64 and RMSE~=~59.73 (  {degree}), 
 and hence a lower predictability of R2~=~0.61. Such an RMSE led to errors around $\pm$60~  {degrees} 
 in a space of $0 \leq \text{WS} < 360$. The influence of residuals in WD can be seen from the outliers in Figure~\ref{fig:raw-scatter-plots}. The reasons are twofold. First, the data collecting procedure is discrete and constrained, accomplished only by a single sampling a minute from the WD sensor. This implies that the wind may change its direction by moving the vane to a new position just before or soon after reading out the sensor. In this occasion, the collected value may not be representative for that whole minute of observation. The second reason is that the time of reading out the WD sensors is not synchronized between the stations. This means that, if the values sampled by the stations are very different, they may represent wind directions of different moments within the observed minute. Although there were these two possibilities of data inconsistencies, they could not cause more impact than the observed outliers. Thus, we understand that the wind direction measurements had little fluctuation within the observed minutes.

From the analysis we carried out, one can conclude that the weather stations produce different weather data, either by the residual or by the significance of paired $t$-tests. In order to reduce residuals while enabling operations with no statistical difference between the stations, we propose to extend the experimental methodology by applying linear and machine learning regression models in order to calibrate the LCAWS weather parameters. In the next section, we discuss experimental results obtained from such a proposal.

\section{Calibrating Weather Sensors with Linear and Machine Learning-Based Regression~Models}
\label{sec:LCAWS-calibration}

Assuming the PWD results as being the expected values, i.e., ground-truth, we applied regression methods in order to reduce the LCAWS residuals. As described in Table~\ref{tab:reg-models-literature}, other works applied regression methods to accomplish sensor calibration or data inference. The authors observed promising results from well-known linear methods such as Linear Regression (LM), Multiple Linear Regression (MLR), as well as from well-known machine learning (ML) based techniques such as Support Vector Machines (SVM), Random Forest (RF), and Neural Networks (NN). Aside from analyzing performance of well-known methods, we also verified the efficiency of sophisticated methods of Ensemble Learning (EL) by combining a set of different base models into super learner meta-models. Although ML-based methods are more computationally costly than the linear models LM and MLR, computing resource constraints in the data correction process should not be of concern. Such a process is a non-real-time task which is carried out after data processing as a Cloud Service (CS) on the server side. 

\begin{table}[h]

\footnotesize
\centering
\caption{Related work on applying regression methods to improve weather sensing process.} 
\label{tab:reg-models-literature}
\setlength{\tabcolsep}{3pt}
\begin{tabular} {c c c    c    c c c c c c }
\toprule & \multicolumn{2}{c }{\textbf{Applications}} && \multicolumn{6}{c}{\textbf{Regression Methods}} \\ 
    \cmidrule{2-3} \cmidrule{5-10}
  \textbf{Reference}   & \textbf{Calibration} & \textbf{Inference} 
                && {\textbf{LM}} & {\textbf{MLR}}  & {\textbf{NN}} & {\textbf{SVM}} & {\textbf{RF}} & {\textbf{EL}} \\ 
    \midrule 
  \cite{Fang.and.Bate:2017} & $\checkmark$ &  
                && & $\checkmark$ \\
  \cite{Kelley.et.al:2019} & & $\checkmark$ 
                &&  & $\checkmark$ \\
  \cite{Sharma.et.al:2011} & & $\checkmark$   
                &&  &  &  & $\checkmark$\\
  \cite{Zimmerman.et.al:2018} & $\checkmark$ & 
                && & & & & & $\checkmark$ \\
  \cite{Yamamoto.et.al:2017} & $\checkmark$  & 
                && $\checkmark$ & & $\checkmark$ \\
  \cite{Cordero.et.al:2018} & $\checkmark$ & 
                && & $\checkmark$ & $\checkmark$ & $\checkmark$ & $\checkmark$ & \\
  \textbf{  {*LCAWS}} & $\checkmark$ &   
                && $\checkmark$ & $\checkmark$ & $\checkmark$ & $\checkmark$ & $\checkmark$ & $\checkmark$ \\
  \toprule 
\end{tabular}

\begin{tabular}{p{8.5cm}}
  {\textbf{*} The method we proposed in this paper for intelligent sensor calibration and data correction.}
  {Label:}  {Linear Model~(LM), Multiple Linear Regression~(MLR), Support Vector Machines~(SVM), Random Forest~(RF), Neural Networks~(NN), Ensemble Learning~(EL).}
\end{tabular}
\end{table}

\subsection{Machine Learning Pipeline}

\begin{figure}[h]
    \centering
    \includegraphics[scale = 0.7]{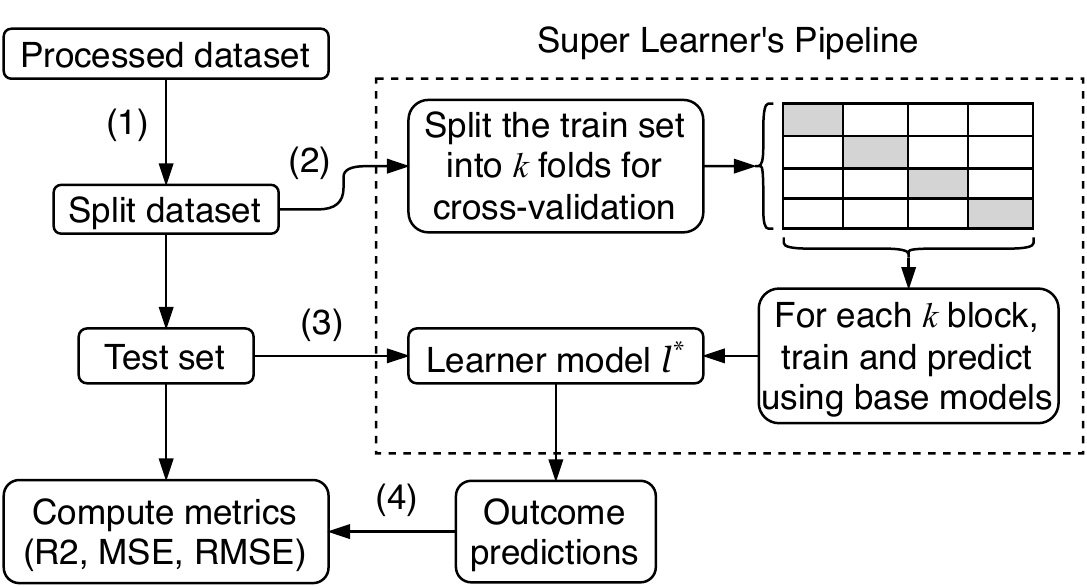}
    \caption{Machine learning regression pipeline we implemented in order to reduce the LCAWS residuals. Four main steps are accomplished: (1)~dataset splitting; (2)~model training and cross-validation; (3)~predictions with the learner model; and (4)~computing performance metrics.
    }
    \label{fig:ML-pipeline}
\end{figure}

In order to reduce the LCAWS residual, we implemented a machine learning pipeline in R language, as illustrated in Figure~\ref{fig:ML-pipeline}. In such a pipeline, there are four important steps. In step~(1), the processed dataset of each sensor is split into two sets. The first set consists of \unit[60]{\%} of the dataset and is used for training and validating the regression models. The second set with the remaining \unit[40]{\%} of the dataset is used for testing the models fitted in step~(1). We observed that the train--test dataset proportion of 60--40{\%} could bring better results, while train sets larger than that have led to overfitting. In step~(2), we submitted the train set to SuperLearner~\cite{Laan.et.al:2007}, which is a machine learning R package~\cite{Package.SuperLearner:2019} for constructing super-learners in predictions based on $k$-fold cross-validations from a set of candidate models. We set $k = 10$~folds in our experiments. Then, learner $l^\star$ is the model or the set of weighted models that most minimizes the risk metric MSE. In step~(3), the fitted model $l^\star$ is then used to make out-of-fold predictions from the test set with the \unit[40]{\%} of the input dataset. Finally, in step~(4), we apply the performance metrics of R2, MSE, and RMSE in order to compare the outcomes $\hat{y}$ predicted by the $l^\star$ model with the expected values $y$ in the test set.

\subsection{Efficiency of the Regression Models}

Considering the processed dataset for the sensors $S = \{$\text{AP}, \text{AT}, \text{RH}, \text{RG}, \text{WS}, \text{WD}$\}$, we evaluated the efficiency of a set of regression models $L = \{$\text{LM}, \text{MLR}, \text{NN}, \text{SVM}, \text{RF}, \text{EL}$\}$ from an amount of different experiments $|E| = \text{100}$, as described in Algorithm~\ref{alg:ML-experiments}. We aim at identifying the best candidate regression models for accomplishing data correction on the weather parameters. Thus, improvements on machine learning methods are beyond the scope of this paper.

\newcommand{\algMLExperiment}{%
    \singlespacing
    \begingroup
    \removelatexerror
    \footnotesize
        \begin{algorithm*}[H]
            \label{alg:ML-experiments}
            \SetKwProg{proc}{procedure}{:}{}
            \SetKwProg{prog}{program}{:}{}
            \DontPrintSemicolon
            \caption{ML experiments.}
            \KwIn{\\
            ~~~~~~$S$ = \{AP, AT, RH, RG, WS, WD\}, the set of sensors. \\
            ~~~~~~$L$ = \{LM, MLR, NN, SVM, RF, EL\}, the set of learning techniques. \\
            ~~~~~~$E$ = \{1, 2, $\times$, 100\}, set of experiments. \\
            ~~~~~~$\overline{\text{DB}}$, processed dataset containing the weather parameters.
            }
            \KwOut{\\
            ~~~~~~$O$, resulting set of multiple ML outcomes, $o = \{l^\star, d, \hat{y}, m\}$.
            }
            \prog{\upshape{\code{run\_experiments($S$, $L$, $E$, $\overline{\text{DB}}$)}}}
            {
                $O$ = \texttt{null}\;
                \ForEach{$s \in S$}{ 
                    $\overline{\text{D}}_s$ = \code{subset($\overline{\text{DB}}$, $s$)}\; 
                    \ForEach{$l \in L$}{
                        \ForEach{$e \in E$} {
                            $o$ = \code{learning\_pipeline($\overline{\text{D}}_s$, $l$, $e$)}\;
                            \code{append($O$, $o$)}
                        }
                    }                
                }
                \KwRet $O$
            }
            \proc{\upshape{\code{learning\_pipeline($\overline{\text{D}}_s$, $l$, $e$)}}}
            {
                \code{set\_seed($e$)}\;
                $d$ = \code{split\_dataset($\overline{\text{D}}_s$, 60\%, 'rand')}\;
                $l^\star$ = \code{super\_learner($d_{\text{train}[y]}$, $d_{\text{train}[x]}$, $l$)}\;
                $\hat{y}$ = \code{predict($l^\star$, $d_{\text{test}[x]}$})\;
                $m$ = \code{compute\_metrics($\hat{y}$, $d_{\text{test}[y]}$)}\;
                \KwRet $o$ = \{$l^\star$, $d$, $\hat{y}$, $m$\}\;
            }  
\end{algorithm*}
\endgroup
}

\begin{center}
    \begin{minipage}[c][10.5cm][t]{8.5cm}
        \algMLExperiment    
    \end{minipage}
\end{center}

In the experiments, we set three models of feed-forward neural networks regarding different sizes for the hidden layer (4, 7, and 10 units). The SVM model was configured with radial basis function and regression precision (nu) of 0.5. The Random Forest (RF) model was parameterized with 1000 trees grown. We defined a set of more than twenty different candidate models\footnote{Candidate models applied for the ensemble learning (EL) from a number of R packages: \texttt{biglasso}, \texttt{extraTrees}, \texttt{gam}, \texttt{glm}, \texttt{glm.interaction}, \texttt{ipredbagg}, \texttt{kernelKnn}, \texttt{ksvm}, \texttt{lm}, \texttt{loess}, \texttt{mean}, \texttt{nnet}, \texttt{nnls}, \texttt{polymars}, \texttt{randomForest}, \texttt{ranger}, \texttt{rpart}, \texttt{rpartPrune}, \texttt{speedglm}, \texttt{speedlm},	 \texttt{step}, \texttt{step.forward}, \texttt{step.interaction}, \texttt{stepAIC}, \texttt{svm}.} for ensemble learning (EL).

From the database \code{DB} with the processed parameters of both LCAWS and PWS stations, for each sensor $s \in S$, we selected the respective processed dataset $\overline{\text{DB}}_s$ and conducted 100~experiments from it. In each experiment $e \in E$, the four steps showing in Figure~\ref{fig:ML-pipeline} are accomplished by the \codeSmall{  {learning\_pipeline}} procedure. Since a random seed is set for each experiment, different random samples of each weather parameters are generated and split by keeping the 60--40{\%} train--test proportion. When having the fitted model $l^\star$ from the SuperLearner's pipeline for training and cross-validation, the predictions are accomplished from the test set and performance metrics are computed. The output $o$ of each experiment consists of the model $l^\star$, the dataset~$d$, the predicted values~$\hat{y}$, and the obtained performance metrics~$m$. Regarding the outputs $o \in O$, we applied $t$-tests to verify the significance of the top-1 fittest model's R2 coefficient against the other models. Thus, we identified whether the best regression model was really able to perform differently than the others.

Table~\ref{tab:ml-ranking} presents the performance ranking based on the average (Avg.) and standard deviations (SD) of the regression models' R2 coefficients, as well as the the paired $t$-tests. The baseline model RAW (highlighted in blue) presents the LCAWS performance results regarding its originally processed weather parameter. Except for RG parameters, the experimental results show significant improvement on the data correctness. When minimizing the residuals, the top-1 regression models allowed for maximizing the performance in R2 with statistical significance in relation to the RAW baseline. Considering R2 and the respective \textit{p}-values for AP, AT, and WS sensors, one can verify that there were no significant differences in the performance of simple regression models (LM, MLR) and more sophisticated ML-based models (EL, SVM, RF). The EL model was the best one for the RH sensor, reaching \textit{p}-value with high significance, differentiating it from the others. For the WD parameter, the ML-based models EL and RF have performance with no statistical difference. For the RG sensor, it is irrelevant applying or not with regression models to correct the RG parameter. This can be verified from the R2 performance of the fittest model LM, which had no significant difference to the ML, RAW (i.e., LCAWS raw data), and EL~models.

\begin{table}[]
\footnotesize
\centering
\caption{Ranking of different regression models and R2 significance (Top-1 model vs. others) from results of 100~experiments.} 
\label{tab:ml-ranking}
\begin{tabular}{ l l l   r l r l}
 \toprule \textbf{Sensor} & \textbf{Model} & \textbf{Rank} & \multicolumn{2}{c}{\textbf{R2}} & \multicolumn{2}{c}{\textbf{R2 Significance}}  \\
     &  &  & \textbf{Avg.} & \textbf{($\pm$SD)} & \textbf{\textit{t}-value} & \textbf{\textit{p}-value} \\
\midrule AP & MLR &     1 & 0.9991 & ($\pm$ 1e-04) &  &   \\ 
   & EL &     2 & 0.9987 & ($\pm$ 0.0025) & 1.62 & 0.1078 \\ \cmidrule{3-7}
   & LM &     3 & 0.9966 & ($\pm$ 3e-04) & 74.85 & **  \\ 
   & SVM &     4 & 0.9871 & ($\pm$ 0.0047) & -25.69 & ** \\ 
   & RF &     5 & 0.9613 & ($\pm$ 0.0059) & -64.57 & ** \\ 
   & {\color{blue}{RAW}} &     6 & 0.9555 & ($\pm$ 0.0035) & 125.61 & **  \\ 
   & NN.4 &     7 & 0.0127 & ($\pm$ 0.142) & 69.47 & ** \\ 
  \midrule AT & MLR &     1 & 0.9939 & ($\pm$ 0.0012) &  &   \\ 
   & EL &     2 & 0.9939 & ($\pm$ 0.0045) & 0.11 & 0.9141 \\ \cmidrule{3-7}
   & LM &     3 & 0.9929 & ($\pm$ 0.0014) & 5.68 & ** \\ 
   & SVM &     4 & 0.9852 & ($\pm$ 0.0046) & -18.31 & ** \\ 
   & RF &     5 & 0.9835 & ($\pm$ 0.0022) & -41.08 & ** \\ 
   & {\color{blue}{RAW}} &     6 & 0.9254 & ($\pm$ 0.0058) & 115.06 & ** \\ 
   & NN.4 &     7 & 0.0340 & ($\pm$ 0.1971) & 48.7 & ** \\ 
  \midrule RH & EL &     1 & 0.9943 & ($\pm$ 0.0049) &  &  \\ \cmidrule{3-7}
   & MLR &     2 & 0.9924 & ($\pm$ 0.0014) & -3.77 & ** \\ 
   & LM &     3 & 0.9910 & ($\pm$ 0.0017) & -6.44 & ** \\ 
   & SVM &     4 & 0.9838 & ($\pm$ 0.0063) & -13.17 & **  \\ 
   & RF &     5 & 0.9826 & ($\pm$ 0.0025) & -21.16 & **  \\ 
   & {\color{blue}{RAW}} &     6 & 0.9177 & ($\pm$ 0.0065) & 94.29 & ** \\ 
   & NN.10 &     7 & 0.0110 & ($\pm$ 0.1206) & -81.44 & **  \\ 
  \midrule RG & LM &     1 & 0.9235 & ($\pm$ 0.1145) &  &  \\ 
   & MLR &     2 & 0.9228 & ($\pm$ 0.1155) & -0.05 & 0.964 \\ 
   & {\color{blue}{RAW}} &     3 & 0.9060 & ($\pm$ 0.0982) & 1.16 & 0.2485  \\ 
   & EL &     4 & 0.8967 & ($\pm$ 0.1066) & 1.71 & 0.0887 \\ \cmidrule{3-7}
   & RF &     5 & 0.4902 & ($\pm$ 0.1446) & -23.5 & **  \\ 
   & SVM &     6 & 0.2220 & ($\pm$ 0.2002) & -30.42 & **  \\ 
   & NN.10 &     7 & 0.0316 & ($\pm$ 0.1891) & 40.34 & **  \\ 
  \midrule WS & EL &     1 & 0.9685 & ($\pm$ 0.0736) &  &  \\ 
   & MLR &     2 & 0.9668 & ($\pm$ 0.0036) & -0.24 & 0.8138  \\ 
   & SVM &     3 & 0.9563 & ($\pm$ 0.0113) & -1.65 & 0.1026  \\ 
   & LM &     4 & 0.9554 & ($\pm$ 0.0051) & -1.78 & 0.0776  \\ \cmidrule{3-7}
   & RF &     5 & 0.9458 & ($\pm$ 0.0089) & -3.06 & *  \\ 
   & {\color{blue}{RAW}} &     6 & 0.3395 & ($\pm$ 0.0213) & 82.1 & **  \\ 
   & NN.10 &     7 & 0.1704 & ($\pm$ 0.3764) & -20.81 & **  \\ 
  \midrule WD & EL &     1 & 0.7231 & ($\pm$ 0.0486) &  &  \\ 
   & RF &     2 & 0.7213 & ($\pm$ 0.037) & -0.29 & 0.7708  \\ \cmidrule{3-7}
   & SVM &     3 & 0.7007 & ($\pm$ 0.0553) & -3.04 & *  \\ 
   & MLR &     4 & 0.6818 & ($\pm$ 0.0532) & -5.73 & **  \\ 
   & LM &     5 & 0.6779 & ($\pm$ 0.0563) & -6.08 & ** \\ 
   & {\color{blue}{RAW}} &     6 & 0.6045 & ($\pm$ 0.0719) & 13.67 & ** \\ 
   & NN.7 &     7 & 0.0195 & ($\pm$ 0.1252) & -52.39 & ** \\ 
   \toprule 
   \multicolumn{7}{l}{Significance codes: 
   $(**) < \text{0.001}$, 
   $(*) < \text{0.01}$}
\end{tabular}
\end{table}

RG and WD sensors have inherent limitations of analog devices based on magnetic reed switches, either by the device design or by the data sampling procedure. Unlike the other parameters, rainy events and wind speed had no pattern. We expect to address these issues to improve the data correctness in further investigations, e.g., hardware device design, software data collecting procedures, and/or improving regression methods.

\subsection{Correcting Weather Parameters with the Fittest Models}

\begin{figure*}[h]
\centering
    \subfigure[~LCAWS weather parameters processed from the raw data.]{
	    \includegraphics[scale=0.73, trim = 0.2cm 0cm 0cm 0cm]{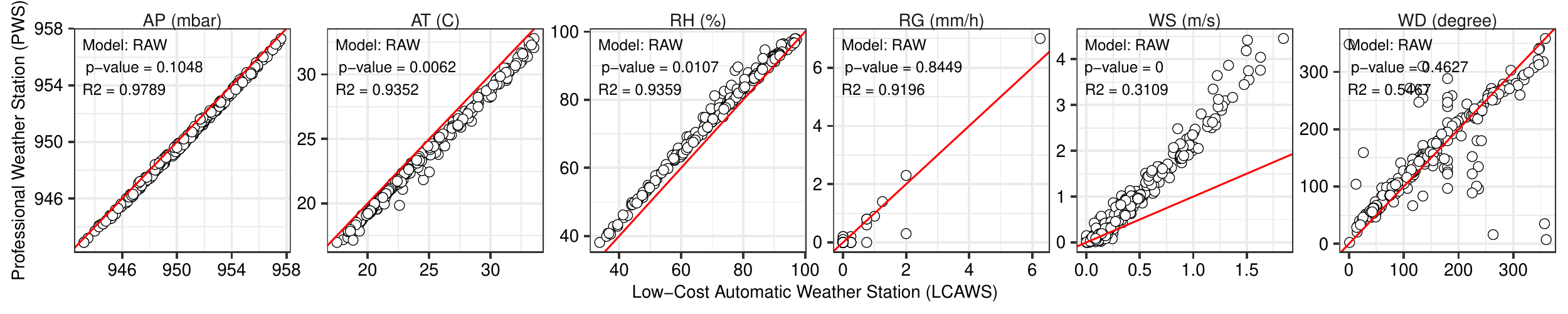}
	    }
	
	\subfigure[~Processed LCAWS weather parameters corrected through the fittest regression models.]{
	    \includegraphics[scale=0.73, trim = 0.2cm 0cm 0cm 0cm]{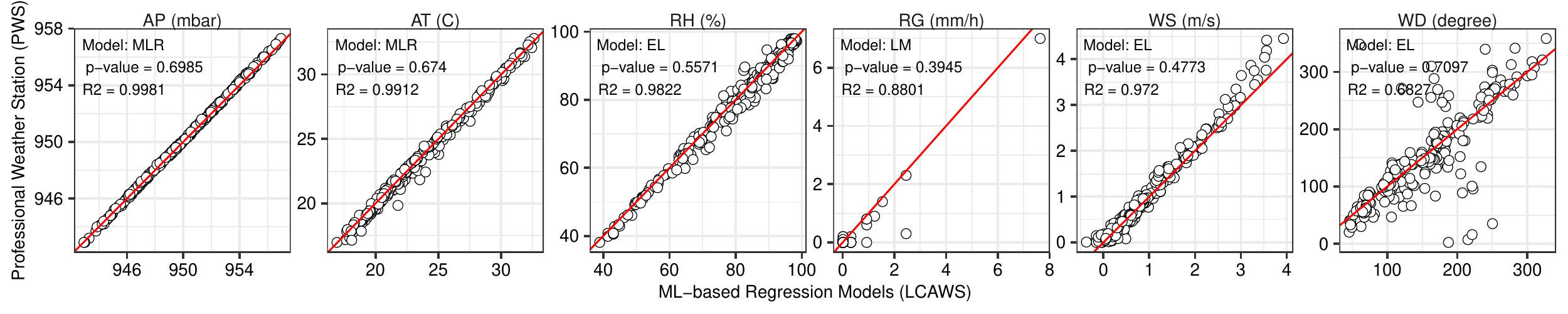}
	    }
	 
\caption{Scatter plots between LCAWS and PWS weather parameters, regarding the test set from day 18 to day 30, with different sensors: Air Pressure~(AP), Air Temperature~(AT), Relative Humidity~(RH), Rain Gauge~(RG), Wind Speed~(WS), and Wind Direction~(WS).}
\label{fig:scatter-tests}
\end{figure*}

Once the best regression models for each sensor are obtained, we carried out a last experiment to verify how data correction would be accomplished in reality. Instead of splitting the processed parameters into randomized samples, we determined the 60--40{\%} train-test proportion with chronological intervals, from day 1  to day 18 and from day 18  to day 30, respectively. Figure~\ref{fig:scatter-tests} shows the scatter plots~(a) and~(b) of the weather parameters with no corrections (RAW data) and the parameters corrected with the fittest regression models, respectively. Table~\ref{tab:ml-stat-test} presents the results from performance metrics. 

\begin{figure}[h]
    \centering
	    \includegraphics[scale=0.73, trim = 0.2cm 0.25cm 0cm 0.4cm]{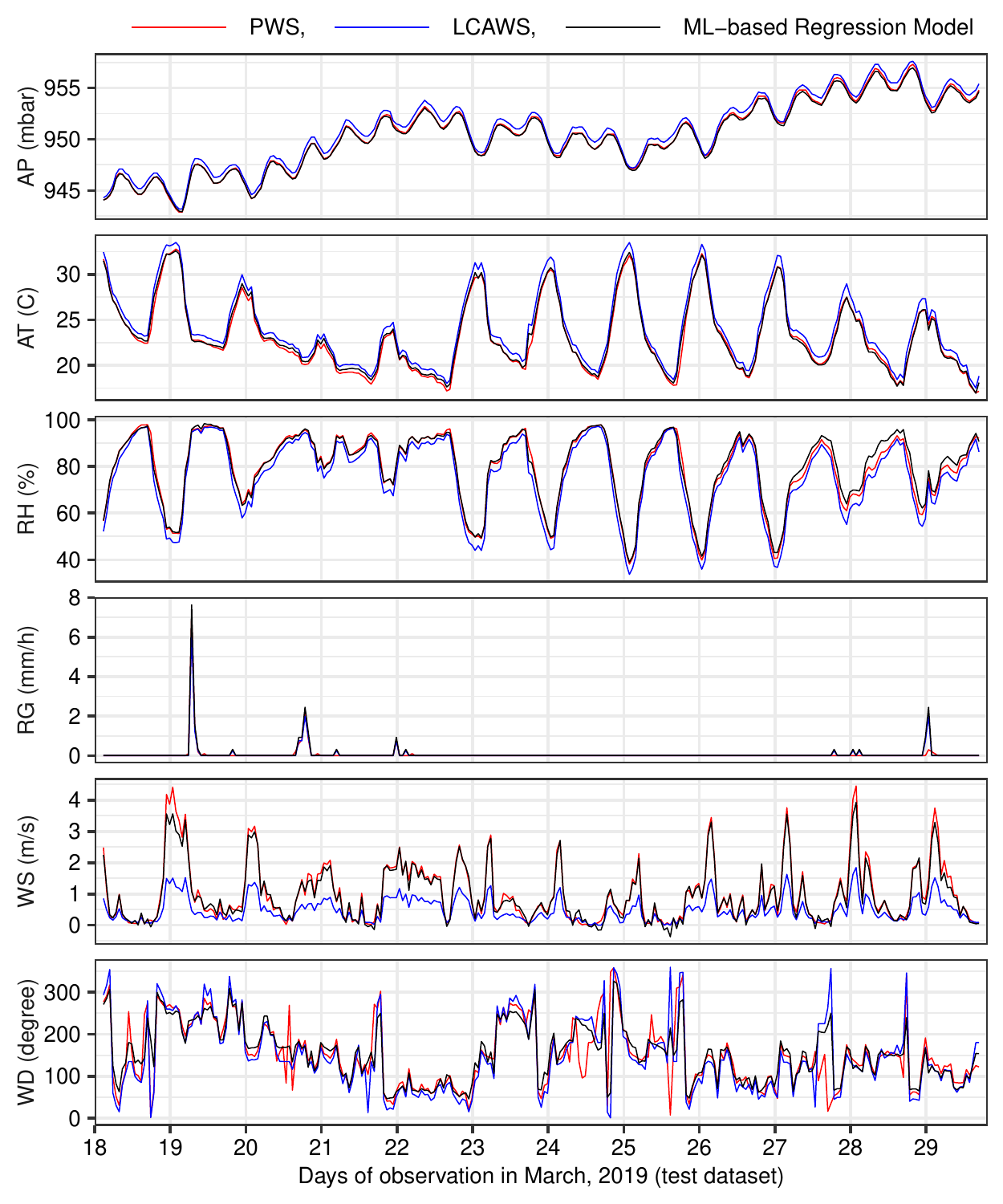}
\caption{Time-series of weather parameters from PWS, LCAWS, regression models, regarding the test set from day 18 to day 30.}
\label{fig:time-series-tests}
\end{figure}

As we previously discussed, we observed that regression models could not reduce the residuals of the RG parameter. In this last experiment, the best model LM in the RG parameters led to a lower R2 with a performance degradation of $-$4.49{\%}. On the other hand, for the other weather parameters, we observed that the LCAWS predicted values $\hat{y}$ from the best regression models could be fitted well to those expected $y$ in PWS. The best models allowed performance improvements in R2 of 1.9{\%}, 5.6{\%}, 6.8{\%}, 68.0{\%}, and 19.9{\%} for the AP, AT, RH, WS, and WD parameters, respectively. In addition, the paired $t$-tests applied in such weather parameters showed \textit{p}-values higher than the level of 0.05, i.e., there was no significant difference between $\hat{y}$ and $y$. These promising fitting results are seen in the time-series in Figure~\ref{fig:time-series-tests}, where the black lines of the $\hat{y}$ values predicted by the best regression models nearly overlap the red line of $y$-values expected in PWS.

\begin{table}[h]
\footnotesize
\centering
\caption{Performance metrics and significance between LCAWS and PWS weather parameters regarding the test set from the day 18 to day 30.} 
\label{tab:ml-stat-test}
\begin{tabular*}{\hsize}{@{}@{\extracolsep{\fill}}lllllrl@{}}
  \toprule \textbf{Sensor} & \textbf{Model} & \textbf{R2} & \textbf{MSE}  & \textbf{RMSE} & \multicolumn{2}{c}{\textbf{Significance}} \\ & & & & & {\textit{\textbf{t}}\textbf{-value}} & {\textit{\textbf{p}}\textbf{-value}} \\ \midrule
AP & RAW & 0.9789 & 0.2316 & 0.4813 & 1.6200 & 0.1048 \\ 
  AT & RAW & 0.9352 & 0.9686 & 0.9842 & 2.7500 & * \\ 
  RH & RAW & 0.9359 & 14.6893 & 3.8327 & $-$2.5600 & $.$ \\ 
  RG & RAW & 0.9196 & 0.0168 & 0.1296 & 0.2000 & 0.8449 \\ 
  WS & RAW & 0.3109 & 0.6404 & 0.8003 & $-$8.6200 & ** \\ 
  WD & RAW & 0.5467 & 2676.4324 & 51.7342 & $-$0.7400 & 0.4627 \\ \midrule
  
  AP & MLR & 0.9981 & 0.0204 & 0.1430 & $-$0.3900 & 0.6985 \\ 
  AT & MLR & 0.9912 & 0.1309 & 0.3618 & 0.4200 & 0.674 \\ 
  RH & EL & 0.9822 & 4.0804 & 2.0200 & 0.5900 & 0.5571 \\ 
  RG & LM & 0.8801 & 0.0251 & 0.1583 & 0.8500 & 0.3945 \\ 
  WS & EL & 0.9720 & 0.0260 & 0.1613 & $-$0.7100 & 0.4773 \\ 
  WD & EL & 0.6827 & 1873.3925 & 43.2827 & 0.3700 & 0.7097 \\ 
   \toprule 
   \multicolumn{7}{l}{Significance codes: 
   (**) $< \text{0.001}$, 
   (*) $< \text{0.01}$, 
   (.) $< \text{0.05}$.}
\end{tabular*}
\end{table}

From the experimental results we obtained, one can conclude that, by applying the regression model-based data correction approach, it was possible to calibrate the sensors AP, AT, RH, WS, and WD. Such an approach allowed: (1)~minimizing the residuals from MSE and RMSE metrics and, hence, maximizing the performance R2 of these LCAWS sensors; (2)~validating statistically these LCAWS sensors with no significant differences of its corrected weather parameters to the expected ones in PWS. With reduced residuals, the \textit{p}-values above 0.05 make the null hypothesis be accepted, i.e., the hypothesis that the weather stations produce equivalent weather parameters. In the case of the RG sensor, whether or not to apply regression models is indifferent. Nevertheless, the original results with processed data from LCAWS show to be statistically similar to the PWS results.

\section{Conclusions}
\label{sec:Conclussions}

Weather stations play a key role for the natural disaster monitoring, being the main source of accurate, reliable, and {in situ} weather data. To support risk management and reduce the impacts of natural disasters in Brazil, Cemaden urges to provide large-scale weather monitoring from an observational network of more than five thousands of data collection platforms (DCPs) spread across risk areas in thousands of cities of the Brazilian territory. The main vulnerabilities of DCPs are not linked to the hardware robustness, but mostly to the applied technology that is provided only from high-cost professional weather instrumentation. While the expensive DCP's components lead to high costs of maintenance for both preventively and replacing failed components, the vulnerabilities in the observational network are mitigated ultimately from short-term periodic maintenance. In this context, the maintenance of DCPs is extremely costly for a vast country with a need for thousands of DCPs such as Cemaden's observational network, mostly under limited human and financial resources.

As an initiative to address cost reduction in maintenance process of natural disaster monitoring, we propose to apply IoT technology based on COTS in order to reduce the cost of weather instrumentation from tens of thousands of dollars to a few hundred. To this end, in this paper, we present a comprehensive material on the design, implementation, validation, and intelligent sensor calibration of a low-cost automatic weather station (LCAWS) entirely developed from commercial-grade parts and open-source IoT technologies. From a robust methodology, we showed that the proposed LCAWS was able to provide weather data as reliably as a professional weather station (PWS) of reference for natural disaster monitoring.





\subsection{Addressing LCAWS's Limitations in Future Work}

As a proof-of-concept, LCAWS is an initial work to demonstrate the feasibility of low-cost alternatives to reduce maintenance costs of the Brazilian weather observational network. However, along the LCAWS validation process, we identified the following hardware and software limitations:

\begin{itemize}
    \item {Datalogger enclosure protection}. The datalogger housing, rated IP55, is not hermetically sealed. Although the silica gel was utilized to keep the interior of the housing dry, small amounts of water condensation were observed on the interior walls of the housing when the silica gel became saturated. It is not known if this water came from the outside or from battery acid evaporation. Moreover, high temperatures were frequently observed inside the datalogger housing during sunny days, up to a maximum of 47 degrees Celsius. Proposed improvements to be investigated are utilizing an enclosure with superior thermal insulation and shielding the enclosure from direct sunlight. 
    
    \item {Short battery life-cycle}. While the power consumption to run RF modules (GPRS, GPS) made it necessary to switch the \unit[10]{Wp} solar panel to a \unit[55]{Wp} one, the battery lasted approximately six months before it could not hold enough charge to last a whole night. This short battery life may be also due to the observed high temperatures inside the housing.
    
    \item {Rain gauge faults}. The original rain gauge design was prone to false pulses when high winds caused the tipping bucket to vibrate. This was successfully corrected by removing the single reed switch, which was triggered when the bucket passed the middle position and replacing it with two reed switches. Each reed switch was then triggered when the bucket sat at the end (tipped)~position. 
    
    \item {AWS' inconsistencies}. 
    A supposed software error causes the Arduino to freeze, requiring a manual hard reset. This problem is recurrent, usually happening once a month, and is still under investigation. While the AWS implementation (Figure~\ref{fig:LCAWS-sysarch}) has validated LCAWS, its coding requires improvements, including refactoring.
\end{itemize}

In this context, further investigations are required to address LCAWS's hardware and software limitations above mentioned and, then, evolve LCAWS to be deployed on Cemaden's observational network and integrated into its natural disaster monitoring~pipeline. 

\section*{Acknowledgments}
This research was partially funded by São Paulo Research Foundation (FAPESP), Grant Nos. \#2015/18808-0, \#2018/23064-8, \#2019/23382-2. 
We thank Antonio Carlos Varela Saraiva (UNESP, Brazil) who allowed the validation of LCAWS to be accomplished together with his research group's professional weather station (PWS) deployed at PqTec~\cite{pqTec:2020}. We also thank Fabio Augusto Faria (UNIFESP, Brazil) for the suggestions about machine-learning best practices that helped us to implement a robust method for intelligent sensor calibration and data correction.

\section*{Data Availability}

Datasets and other complementary materials can be {found in}~\cite{complementary.material:2021}.

\bibliographystyle{elsarticle-harv}
\bibliography{references}

\end{document}